\documentclass{article}
\usepackage{geometry}
\usepackage[utf8]{inputenc}
\usepackage{cite}
\usepackage{url}
\usepackage{amssymb,amsmath,amsthm}
\usepackage{bm}
\usepackage{graphicx}
\usepackage{enumerate}
\usepackage{microtype}
\usepackage{color}
\usepackage{hyperref}
\usepackage{multirow}
\usepackage{array}
\newcolumntype{C}[1]{>{\centering\let\newline\\\arraybackslash\hspace{0pt}}m{#1}}

\usepackage{algorithm}
\usepackage{algpseudocode}
\usepackage{amsmath}
\usepackage{amssymb}
\newcolumntype{C}[1]{>{\centering\let\newline\\\arraybackslash\hspace{0pt}}m{#1}}

\newtheorem{defn}{\noindent $\mathbf{Definition}$}[section]

\newtheorem{proposition}[defn]{$\mathbf{Proposition}$}
\newtheorem{theorem}[defn]{$\mathbf{Theorem}$}

\newtheorem{remark}[defn]{$\mathbf{Remark}$}

\title{Volumetric Parameterization for 3-Dimensional Simply-Connected Manifolds}

\author{Zhiyuan Lyu\thanks{Department of Mathematics, The Chinese University of Hong Kong
  ({zylyu@math.cuhk.edu.hk}).}
\and Qiguang Chen\thanks{Department of Mathematics, The Chinese University of Hong Kong
  ({qgchen@math.cuhk.edu.hk}).}
\and Gary P. T. Choi\thanks{Department of Mathematics, The Chinese University of Hong Kong
  ({ptchoi@cuhk.edu.hk}).}
\and Lok Ming Lui\thanks{Department of Mathematics, The Chinese University of Hong Kong
  ({lmlui@math.cuhk.edu.hk}).}}

\date{}

\begin{document}

\maketitle
\begin{abstract}
With advances in technology, there has been growing interest in developing effective mapping methods for 3-dimensional objects in recent years. Volumetric parameterization for 3D solid manifolds plays an important role in processing 3D data. However, the conventional approaches cannot control the bijectivity and local geometric distortions of the result mappings due to the complex structure of the solid manifolds. Moreover, prior methods mainly focus on one property instead of balancing different properties during the mapping process. In this paper, we propose several novel methods for computing volumetric parameterizations for 3D simply-connected manifolds. Analogous to surface parameterization, our framework incorporates several models designed to preserve geometric structure, achieve density equalization, and optimally balance geometric and density distortions. With these methods, various 3D manifold parameterizations with different desired properties can be achieved. These methods are tested on different examples and manifold remeshing applications, demonstrating their effectiveness and accuracy.
\end{abstract}

\section{Introduction}

In recent years, 3D data has become increasingly important in various fields, including computer graphics, computer vision, and medical imaging. Unlike 2D data, 3D data exhibits more complex and irregular structures, making it more challenging to process and manipulate. One of the most important techniques to address these challenges is parameterization, mapping complex manifolds onto canonical domains while preserving desired properties. In particular, surface parameterizations have been widely applied in processing 3D data.  

Various surface parameterization methods have been proposed depending on the specific properties required for different applications. There are two major classes for surface parameterization: conformal parameterization and authalic parameterization. Conformal parameterization preserves the angles and geometry locally while ignoring the area distortions~\cite{choi2015flash}. By contrast, authalic parameterization maintains area measure and neglects the geometric errors~\cite{su2016area}. 

3D manifold parameterization has recently attracted increasing attention as a powerful technique for mapping volumetric domains into a standardized reference space. Compared with surfaces, 3D solid manifolds contain interior volume structure, making their parameterization more difficult. Various methods have been proposed to address those challenges~\cite{yueh2019novel,tan2025dimensional,zhang2022unifying,su2017volume}. However, there remain some gaps for the 3D manifold parameterization. Current approaches suffer several critical limitations. These methods haven't considered geometric information during the mapping process. This shortage leads to the result parameterizations containing large geometric distortions. Additionally, existing methods have difficulty maintaining bijectivity, which is a fundamental requirement for some applications. These limitations pose significant obstacles to the application of current methods in the real world.

Motivated by this, we propose in this paper several novel algorithms for computing the 3D volumetric parameterization for simply-connected 3D manifolds, which satisfy different properties. Unlike the prior methods, our proposed work considers both geometric and authalic distortions. Here, we propose three models: First, we develop a novel method for computing bijective 3D quasi-conformal maps (abbreviated 3DQC) by minimizing the energy formula involving geometric terms. Then we present a method for computing bijective 3D density-equalizing maps (abbreviated 3DDEM) utilizing the diffusion process. Finally, by combining the density-equalization and 3D quasi-conformality, we propose an energy minimization model for producing the bijective 3D density-equalizing quasi-conformal maps (abbreviated 3DDEQ) that achieve an optimal balance between the geometric and volumetric distortions. The experimental results and practical applications demonstrate the effectiveness of our proposed framework.

The contributions of this paper can be summarized as follows:
\begin{enumerate}
\item We present novel approaches for computing parameterizations of simply-connected 3D manifolds that overcome the bijectivity and geometric limitations of conventional methods.
\item Our approach efficiently minimizes local geometric distortion while guaranteeing bijective mappings.
\item Our algorithm is capable of computing a bijective density-equalizing parameterization for simply-connected 3D manifolds, which controls the volumetric distortion efficiently. 
\item Our algorithm is the first parameterization approach that incorporates 3D quasi-conformal techniques to minimize local geometric distortions in density-equalizing parameterizations
\item Our proposed algorithms can be effectively applied to manifold remeshing.
\end{enumerate}

The organization of the paper is as follows. In Section \ref{sec:related_work}, we review the previous works on surface and manifold parameterization. In Section \ref{sec:background}, we introduce the mathematical concepts related to our work. In Section~\ref{sec:proposed_model}, we present our proposed mathematical models for computing bijective 3D manifold parameterizations. In Section \ref{sec:main}, we describe the details of our proposed algorithms. Experimental results are presented in Section \ref{sec:experiment} to demonstrate the effectiveness of our proposed method. In Section \ref{sec:applications}, we present applications of our method in different fields. We conclude our work and discuss future directions in Section \ref{sec:discussion}.

\section{Related work}\label{sec:related_work}
In this section, we briefly review some previous works closely related to this paper.

Surface parameterization has been widely studied in computer graphics, geometry processing, and imaging science. For surface parameterization, there are two major classes for surface parameterization: conformal parameterization and authalic parameterization. Conformal parameterization preserves the angles and geometry locally while ignoring the area distortions. By contrast, authalic parameterization maintains area measure and neglects the geometric errors. For conformal parameterization, numerous methods have been established, including: linearization of the Laplace equation~\cite{angenent1999conformal}, least-square conformal mapping (LSCM)~\cite{levy2023least}, Dirichlet energy minimization~\cite{gu2004genus}, Yamabe flow~\cite{luo2004combinatorial,zeng2012computing}, parallelizable global conformal parameterization~\cite{choi2020parallelizable}, fast disk conformal map~\cite{choi2024fast}, and quasi-conformal composition~\cite{choi2015flash,choi2016spherical,choi2018linear}. 
Compared to the conformal case, there are fewer authalic parameterization methods. Proposed methods includes locally authalic map~\cite{nadeem2016spherical}, optimal mass transport~\cite{zhao2013area,su2016area}, Liesterth energy minimization~\cite{yueh2019novel},
and density-equalizing map~\cite{gastner2004diffusion,gastner2018fast,choi2018density,lyu2024bijective,lyu2024spherical,lyu2024ellipsoidal}.

Parameterization of 3D manifolds is more challenging than the surface case due to their complex structures. In recent years, there has been an increasing interest in volumetric parameterization methods. For instance,  Paill´e and Poulin~\cite{paille2012conformal}, and Chern et al.~\cite{chern2015close} proposed conformal volumetric mappings based on the Cauchy-Riemann equation on each canonical orthogonal plane in $\mathbb R^3$. Kovalsky et al.~\cite{kovalsky2015large} introduced a method for computing volumetric parameterization with large-scale bounded distortion. Rabinovic et al.~\cite{rabinovich2017scalable} developed a flip-preventing mapping for 3D mesh parameterization by minimizing a linear combination of local isometric distortions. Jin et al.~\cite{jin2015stretch} proposed a method for volumetric parameterization by minimizing the stretch-distortion energy. Yueh et al.~\cite{yueh2019novel,yueh2020new} proposed volumetric stretch energy minimization methods for computing the 3D manifolds with different topologies. Su et al.~\cite{su2017volume} developed a volume-preserving parameterization method for tetrahedral meshes based on optimal mass transport theory.

\section{Mathematical background}\label{sec:background}
In this section, we describe some basic theories of density-equalizing maps and quasi-conformal geometry.

\subsection{Density-equalizing maps}\label{sse:density-equalizing}
\begin{figure}[t]
   \centering
   \includegraphics[width=\textwidth]{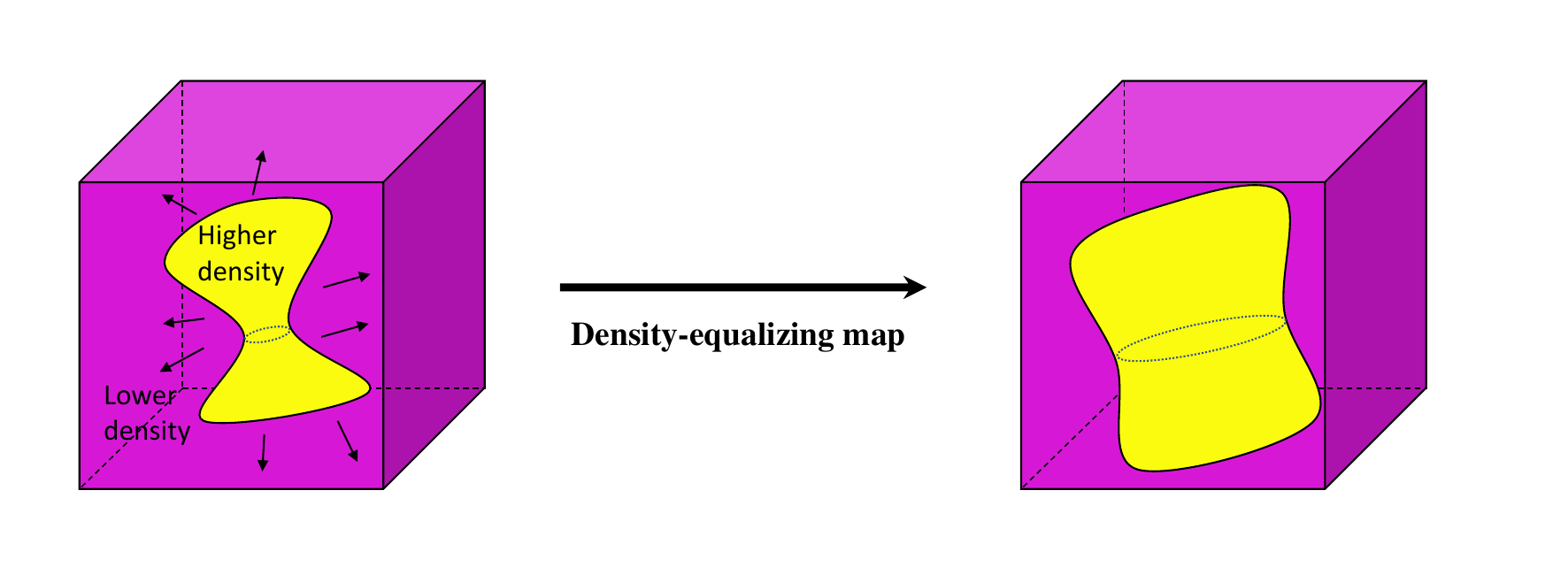}
    \caption{\textbf{An illustration of density-equalizing maps.} During the diffusion process, the regions with high density will be enlarged and the regions with low density will be shrunk. }
    \label{fig:DEM_illustration}
\end{figure}

The goal of \emph{density-equalizing maps} is to generate a map that deforms the initial domain to a target domain such that the density of this domain is equalized. This goal can be achieved by deforming the domain in a way that enlarges the regions with high density and shrinks the regions with low density.
Gastner and Newman~\cite{gastner2004diffusion} proposed a method for producing area cartograms based on the diffusion process. In a given planar domain, a positive density function $\rho$ is initially defined on each unit area. 
The map can be deformed  by the advection equation
\begin{equation}\label{advection}
    \frac{\partial \rho}{\partial t} = - \nabla \cdot \mathbf{j},
\end{equation}
where $\mathbf{j} = - \nabla \rho$ is the flux by Fick's law of diffusion. This yields the diffusion equation
\begin{equation}\label{eqt:duffusion}
    \frac{\partial \rho}{\partial t} = \Delta \rho.
\end{equation}
By the definition of flux, the expression of the velocity field in terms of density is 
\begin{equation}\label{eqt:velocity}
    \mathbf{v} = \frac{\mathbf{j}}{\rho} = -\frac{\nabla \rho}{\rho}.
\end{equation}
By integrating the velocity field, the displacement of the tracers $\mathbf{x}(t)$ of any point on the map at time $t$ can be calculated:
\begin{equation}\label{displacement}
    \mathbf{x}(t) = \mathbf{x}(0) + \int^{t}_{0} \mathbf{v}(\mathbf{x},\tau) \mathrm{d\tau},
\end{equation}
where $\mathbf{x}(0)$ is the initial state. In the limit $t \rightarrow \infty$, the density $\rho$ induced by the above displacement $\mathbf{x}(t)$ will be fully equalized per unit area. An illustration is given in Fig.~\ref{fig:DEM_illustration}.


\subsection{2D quasi-conformal theory}\label{ssec:quasi-conformal}
\begin{figure}[t]
   \centering
   \includegraphics[width=\textwidth]{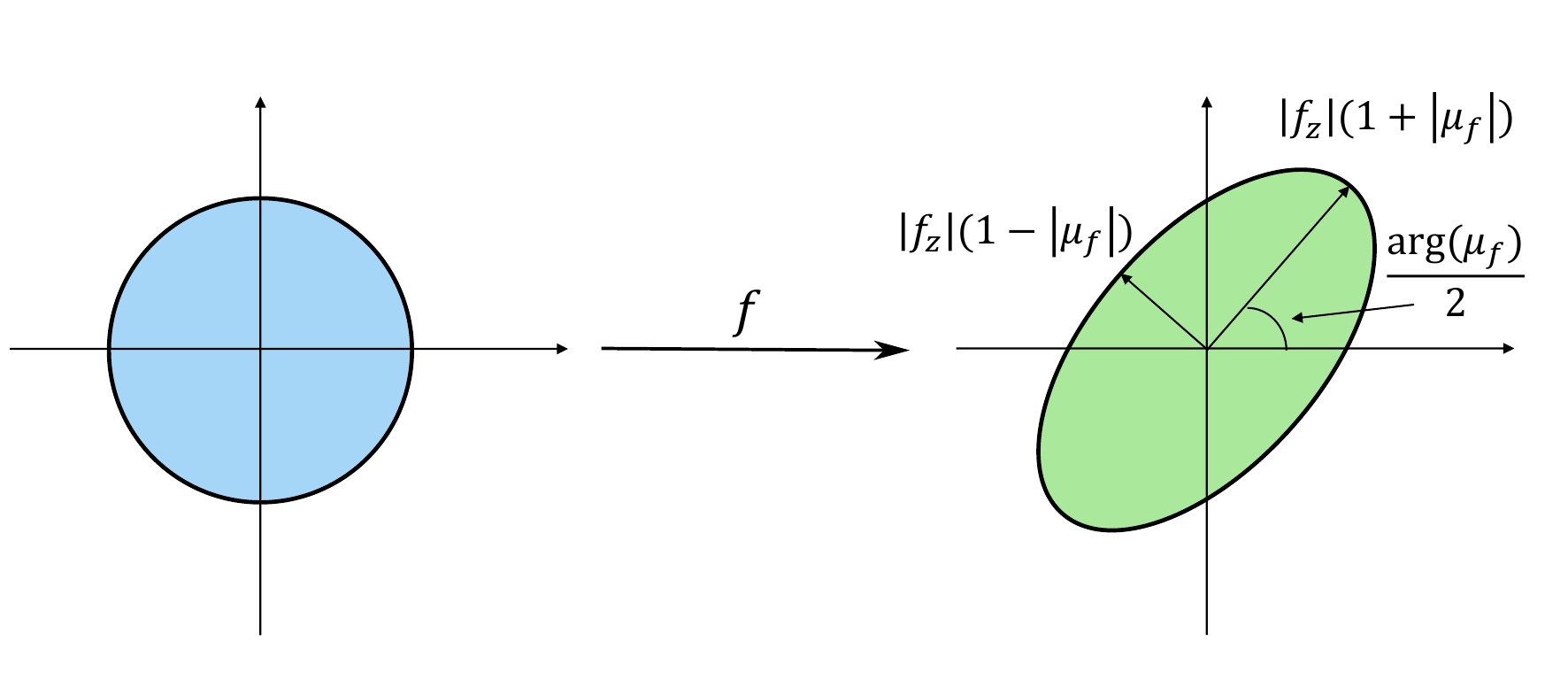}
    \caption{\textbf{An illustration of quasi-conformal maps.} Under a quasi-conformal map $f$, an infinitesimal circle is mapped to an infinitesimal ellipse with bounded eccentricity. The maximal magnification, maximal shrinkage, and maximal dilation are all related to the Beltrami coefficient $\mu_f$.}
    \label{fig:quasiconformal_map}
\end{figure}

\textit{Quasi-conformal maps} is a generalization of conformal maps, which allows some bounded angle distortion. Mathematically, an orientation-preserving homeomorphism $f:\overline{\mathbb C} \rightarrow \overline{\mathbb C}$ is called a quasi-conformal map if it satisfies the Beltrami equation
\begin{equation}\label{eqt:Beltrami_eq}
    \frac{\partial f}{\partial \Bar{z}} = \mu_f(z) \frac{\partial f}{\partial z}
\end{equation}
for some complex-valued function $\mu_f$ satisfying $\|\mu_f \|_{\infty}<1$. $\mu$ is called the \textit{Beltrami coefficient} of $f$, which measures the conformality distortion of $f$. In particular, if $\mu_f = 0$, then Eq.~\eqref{eqt:Beltrami_eq} becomes the Cauchy--Riemann equation, and hence the mapping $f$ is conformal. Intuitively, around a point $z_0 \in \mathbb C$, with respect to its local parameter, the first order approximation of $f$ can be expressed as:
\begin{equation}\label{eqt:first_order_approximation}
    f(z) \approx f(z_0) + f_{z}(z_0)(z-z_0) + f_{\Bar{z}}(z_0)\overline{(z-z_0)} = f(z_0) + f_{z}(z_0)(z-z_0 + \mu_f(z_0)\overline{(z-z_0)}).
\end{equation}
The above formula suggests that $f$ can be considered as a map composed of a translation to $f(z_0)$ combined with a stretch map given by $S(z) = z + \mu(z_0) \Bar{z}$, which is postcomposed by a multiplication of $f_{z}(z_0)$, which is conformal. From $\mu(z_0)$, we can determine the angles at which the magnification and shrinkage are maximized, as well as quantify the degree of magnification and shrinkage at those angles (see Fig.~\ref{fig:quasiconformal_map}). More specifically, the angle of maximal magnification is $\operatorname{arg}(\mu_f(z_0))/2$ with the magnifying factor $|f_{z}(z_0)|(1+|\mu_f(z_0)|)$, and the angle of maximal shrinkage is $(\operatorname{arg}(\mu_f(z_0))+\pi)/2$ with the shrinking factor $|f_{z}(z_0)(|1-|\mu_f(z_0)|)$. The maximal dilation of $f$ is $K(f) = \frac{1+\|\mu_f \|_{\infty}}{1-\|\mu_f\|_{\infty}}$. Thus, the Beltrami coefficient encodes important geometric information about the quasi-conformal map $f$.

Meanwhile, the Beltrami coefficient is closely related to the bijectivity of the quasi-conformal map. More specifically, we have the following result~\cite{lehto1973quasiconformal,ahlfors2006lectures}:
\begin{theorem}\label{qc_bijective}
    If $f$ is a $C^1$ mapping satisfying $\|\mu_f \|_{\infty}<1$, then $f$ is bijective. 
 \end{theorem}


Moreover, let $f,g: \overline{\mathbb{C}} \rightarrow \overline{\mathbb{C}}$ be two quasi-conformal maps with Beltrami coefficient $\mu_f$ and $\mu_g$, respectively. The Beltrami coefficient of the composition map $g\circ f$ can be expressed in the following 
\begin{equation}\label{composition_BC}
    \mu_{g \circ f} = \dfrac{\mu_f + (\overline{f_z} / f_z) (\mu_{g}\circ f) }{1 + (\overline{f_z} / f_z)  \overline{\mu_f}(\mu_g \circ f)}.
\end{equation}
In particular, if $g$ is a conformal map, the Beltrami coefficient satisfies $\mu_{g \circ f} = \mu_f$, implying that composing a conformal map with a quasi-conformal map preserves the coefficient. Furthermore, by choosing a map $g$ with $\mu_g = \mu_{f^{-1}}$, the composition map $g\circ f$ eliminates angle distortions( as $\mu_{g\circ f} = 0$), yielding a conformal map.

\subsection{3D quasi-conformal theory}\label{ssec:3d-quasi-conformal}
\begin{figure}[t]
   \centering
   \includegraphics[width=\textwidth]{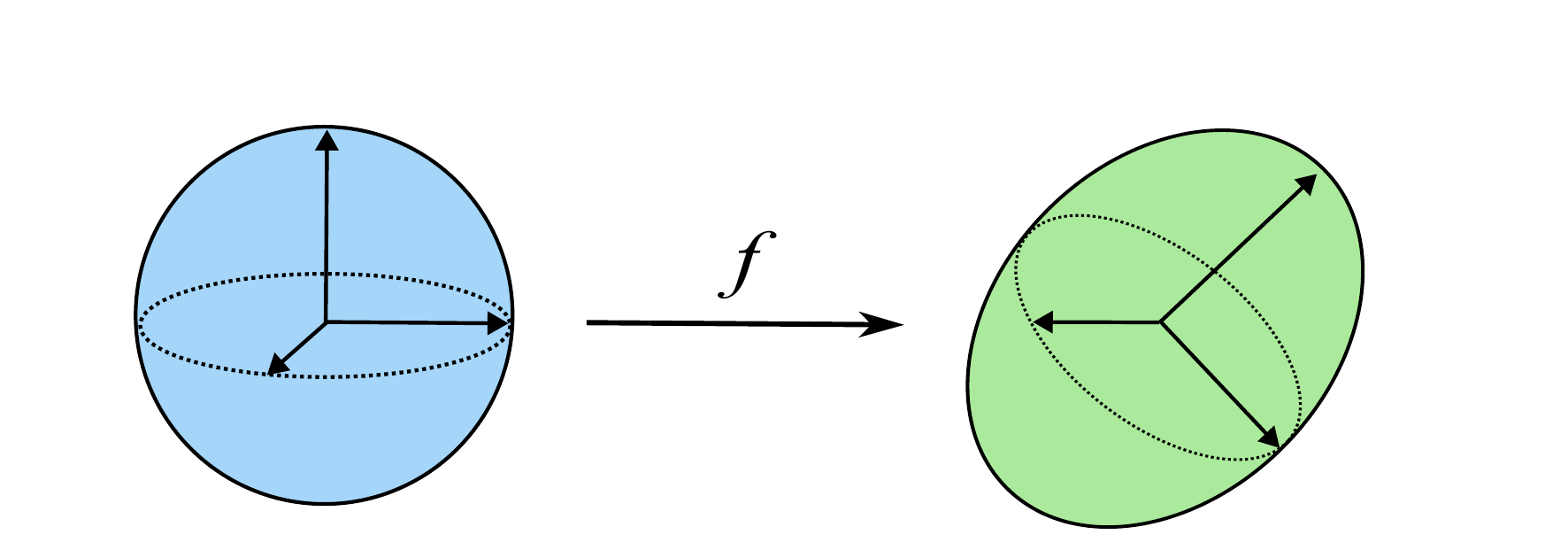}
    \caption{\textbf{An illustration of 3D quasi-conformal maps.} Under a 3D quasi-conformal map $f$, an infinitesimal solid ball is mapped to an infinitesimal solid ellipsoid with bounded eccentricity.}
    \label{fig:3d_quasiconformal_map}
\end{figure}

3D quasi-conformal theory is an extension of the 2D quasi-conformal theory. In~\cite{chen2024novel3dmappingrepresentation}, Chen and Lui proposed a \textit{3D quasi-conformal representation}, together with an effective algorithm for reconstructing a 3D quasi-conformal mapping from its 3D representation. Similar to the Beltrami coefficient, this representation can effectively describe the local dilation of a 
3D mapping (see Fig.~\ref{fig:3d_quasiconformal_map} for an illustration). Moreover, let $f:\Omega_1 \rightarrow \Omega_2$ be a diffeomorphism with $\Omega_1$ and $\Omega_2$ being two open subsets of $\mathbb{R}^3$. The corresponding Jacobian matrix of $f$ is denoted as $\mathbf{J}_f$. By polar decomposition, the $\mathbf{J}_f$ can be divided into:
\begin{equation}\label{eqt:decomposition}
    \mathbf{J}_f = UP,
\end{equation}
where $U$ is an orthogonal matrix and $P = \sqrt{\mathbf{J}^T_f\mathbf{J}_f} = W\Sigma W^{-1}$. Specifically, $\Sigma$ is a diagonal matrix consisting of three positive eigenvalues $\lambda_1$, $\lambda_2$, and $\lambda_3$.

Based on the above formulas, the transpose of the Jacobian matrix $\mathbf{J}^T_f$ can be expressed as:
\begin{equation}
    \mathbf{J}^T_f = W\begin{pmatrix}
    \frac{\lambda_1}{\lambda_2 \lambda_3} & 0 & 0 \\
    0 & \frac{\lambda_2}{\lambda_1 \lambda_3} & 0 \\
    0 & 0 & \frac{\lambda_3}{\lambda_1 \lambda_2}
    \end{pmatrix}W^{-1} \text{Adj}(\mathbf{J}_f),
    \label{eqt:J_f_T}
\end{equation}
where $\text{Adj}(\mathbf{J}_f)$ is the adjugate matrix of $\mathbf{J}_f$, and it takes the form:
\begin{equation}
    \text{Adj}(\mathbf{J}_f) = \begin{pmatrix}
\nabla v \times \nabla w & \nabla w \times \nabla u & \nabla u \times \nabla v
\end{pmatrix}.
\end{equation}
By left multiplying the following matrix on both sides of Eq.~\eqref{eqt:J_f_T},
\begin{equation}
    \mathcal{A} = W\begin{pmatrix}
    \frac{\lambda_2 \lambda_3}{\lambda_1} & 0 & 0 \\
    0 & \frac{\lambda_1 \lambda_3}{\lambda_2} & 0 \\
    0 & 0 & \frac{\lambda_1 \lambda_2}{\lambda_3}
    \end{pmatrix}W^{-1}.
\end{equation}
We obtain the following equation
\begin{equation}
    \mathcal{A}\begin{pmatrix}
\nabla u  & \nabla v & \nabla w 
\end{pmatrix} = \begin{pmatrix}
\nabla v \times \nabla w & \nabla w \times \nabla u & \nabla u \times \nabla v
\end{pmatrix}.
\end{equation}
By applying the divergence operator $\nabla \cdot$ to each column of the transformed equation, we obtain the following system of equations:
\begin{align}
    \nabla \cdot\mathcal{A}\nabla u &= 0, \\
    \nabla \cdot\mathcal{A}\nabla v &= 0, \\
    \nabla \cdot\mathcal{A}\nabla w &= 0.
\end{align}
Finally, these equations can be discretized into a linear system, which allows the 3D mapping $f$ to be computed efficiently.

\section{Proposed mathematical models}\label{sec:proposed_model}
This section presents our proposed framework for 3-dimensional solid manifolds. Below, we first introduce the 3D quasi-conformal coefficient. Then, we propose the variational models focused on different properties.

\subsection{3D quasi-conformal coefficient}
Before introducing our framework, we first review the quasi-conformal theory in the 2D case. As described in Section~\ref{ssec:quasi-conformal}, the relationship between the quasi-conformal map $g$ and its Beltrami coefficient $\mu$ is one-to-one. More specifically, any quasi-conformal map $g$ uniquely determines a measurable Beltrami coefficient $\mu$. On the contrary, the Beltrami coefficient $\mu$ admits a quasi-conformal solution $g$, which is uniquely determined up to post-composition with a conformal map. Furthermore, based on the Beltrami coefficient, we define the dilation $K$ as:
\begin{equation}
    K = \frac{1+ |\mu|}{1- |\mu|}
\end{equation}
to express the geometric deformation of $g$. When $K = 1$, it means the Beltrami coefficient $\mu = 0$, which implies that $g$ is conformal.

In the 2D case, the Beltrami coefficient $\mu$ can clearly reflect the bijectivity of $g$. More specifically, the map $g$ is bijective if and only if its Beltrami coefficient satisfies $\|\mu \|_{\infty} <1$. Analogously, we extend this property into the 3D case.
\begin{theorem}\label{sign_theorem}
    Let $\mathcal{M} \subseteq \mathbb{R}^3$ be a solid domain and $f: \mathcal{M} \to \mathbb{R}^3$ be a $C^1$ map. Under rotation, the Jacobian matrix $J_f$ admits a unique decomposition:
    \begin{equation}
        J_f = UP
    \end{equation}
    where $U$ is a rotation matrix and $P$ is a dilation matrix.
\end{theorem}
\textit{Proof}
When $\operatorname{det}(J_f)\geq 0$, Eq.~\eqref{eqt:decomposition} shows that the Jacobian matrix $J_f$ admits a unique decomposition for all $x\in\mathcal{M}$ via the polar decomposition:
\begin{equation}
    J_f = UP
\end{equation}
where $U$ is an orthogonal matrix and $P = \sqrt{\mathbf{J}^T_f\mathbf{J}_f}$ is a dilation matrix. As $\operatorname{det}(J_f)$ and $\operatorname{det}(P)$ share the same sign, we know that $\operatorname{det}(U)=1$, indicating $U$ is a rotation matrix.

We then consider the case where $\operatorname{det}(J_f)< 0$. Note that Eq.~\eqref{eqt:decomposition} also holds for this case. Since $P = \sqrt{\mathbf{J}^T_f\mathbf{J}_f}$ is a positive definite matrix, $\operatorname{det}(P)>0$. The condition $\operatorname{det}(J_f)< 0$ thus implies $\operatorname{det}(U)< 0$. To proceed, we introduce a transition matrix $L = WDW^{-1}$, where $\operatorname{det}(D)=-1$ and $W$ is the matrix of eigenvectors of $P$. The Jacobian matrix can be rewritten as:
\begin{equation}
    J_f = UP = UL^{-1}LP = \widetilde{U}\widetilde{P},
\end{equation}
where $\widetilde{U} = UL^{-1}$ and $\widetilde{P} = LP$. It is clear that $\operatorname{det}\left(\widetilde{U} \right) = 1$, which implies $\widetilde{U}$ is a rotation matrix. This completes the existence of the desired decomposition.

Next, we provide proof of uniqueness. From the decomposition, the determinant of $J_f$ is:
\begin{equation}
    \operatorname{det}\left(J_f \right) = \operatorname{det}(U)\operatorname{det}(P) = abc,
\end{equation}
where $U$ is a rotation matrix and $P$ is a dilation matrix. $a,b,c$ are the eigenvalues of the dilation matrix $P$. The uniqueness of $\operatorname{det}\left(J_f \right) \geq 0$ is obvious, we only consider $\operatorname{det}\left(J_f \right) < 0$. 

$\operatorname{det}\left(J_f \right) < 0$ holds in two scenarios: 
\begin{itemize}
    \item all eigenvalues negative: $a,b,c<0$;
    \item one negative eigenvalue, two positive: e.g. $a,b >0$ and $c<0$.
\end{itemize}

We show that these cases are equivalent under a rotation.

\textbf{All eigenvalues negative:} We introduce a transition matrix $L = WRW^{-1}$, where $R = \operatorname{diag}(-1, -1, 1)$. Then the Jacobian matrix can be rewritten as:
\begin{equation}
        J_f = UL^{-1}LP = \widetilde{U}LP =\widetilde{U}\widetilde{P},
\end{equation}
where $\widetilde{U} = UL^{-1}$ and $\widetilde{P} = W\widetilde{\Sigma }W^{-1} = WR\Sigma  W^{-1}$. The eigenvalues of $\widetilde{P}$ are $\left(\widetilde{a} ,\widetilde{b},\widetilde{c}\right) = \left(-a,-b,c \right)$. Since $a,b,c<0$, $\widetilde{P}$ has two positive eigenvalues $\widetilde{a},\widetilde{b} >0$ and one negative $c<0$. 

\textbf{One negative eigenvalue, two positive:} The same construction applies. For example, if $P$ has eigenvalues $a<0$ and $b,c >0$, choosing transition matrices $L = WRW^{=1}$ with $R = \operatorname{diag}(-1,1,-1)$ flips the sign of $a$ and $c$, yielding eigenvalues $(-a,b,-c)$.

In both cases, we have $\widetilde{P} = \widetilde{U}^{-1}UP$. Since $\widetilde{U}^{-1}U$ is also a rotation matrix, we can conclude that $\widetilde{P}$ and $P$ are equivalent to each other up to rotation.  Therefore, the decomposition is unique under a rotation.


\hfill $\blacksquare$ 

\begin{remark}
Based on the theorem~\ref{sign_theorem}, we can establish the relation between $\operatorname{det}\left(J_f \right)$ and the eigenvalues $a,b,c$ as follows:
    \begin{itemize}
        \item $\operatorname{det}\left(J_f \right) >0 \Leftrightarrow a,b,c >0$(all positive eigenvalues);
        \item $\operatorname{det}\left(J_f \right) <0 \Leftrightarrow a,b>0, c<0$(two positive eigenvalue, one negative).
    \end{itemize}
\end{remark}

Inspired by the 2D quasi-conformal dilation, the 3D quasi-conformal coefficient is defined by:
\begin{defn}[3D Quasi-Conformal Coefficient]\label{def:3D_coefficient}
    Let $f: \Omega \subset \mathbb{R}^3 \to \mathbb{R}^3$ be a $C^1$ map. The \emph{3D quasi-conformal coefficient} $K$ of $f$ is defined as:
    \begin{equation}\label{eqt:K}
    K(a,b,c) = \frac{\text{max}(a,b,c)}{\text{min}(a,b,c)} = \frac{a}{c},
    \end{equation}
    where $a, b, c$ are the eigenvalues of the dilation matrix induced by $f$ satisfying $a \geq b \geq c$.
\end{defn}

\begin{remark}
      Notably, $K(a, b, c) = 1$ if and only if $a = b = c$, indicating the Jacobian matrix of $f$ has three equal eigenvalues. This condition implies that the mapping $f$ results solely in a rotation. Moreover, $K(f)$ quantifies the local geometric distortion of $f$, with larger values indicating greater geometric distortion.
\end{remark}


Based on the above theorem, the relationship between $K$ and bijectivity can be explained as:
\begin{proposition}\label{bijective_proposition}
    If $f:\mathcal{M} \rightarrow \mathbb B$ is a $C^1$ map satisfying $K$ is positive, then, $f$ is bijective.
\end{proposition}
It's easy to see that $K\geq 1$ when $f$ is a bijective map. 

Similar to the Beltrami coefficient $\mu$, the 3D quasi-conformal coefficient  $K$ reflects geometric distortions. When $K$ is close to 1, the geometric distortions of the mapping $f$ are minimal. Conversely, as $K$ increases, the geometric distortions of $f$ also increase. 
Thus, we can manage the geometric errors on 3D manifolds by controlling the corresponding 3D quasi-conformal coefficient $K$.

In the following, we propose a framework for volumetric parameterization of 3D solid manifolds. First, we present a 3D quasi-conformal method based on the quasi-conformal theory. Next, we introduce a bijective density-equalizing model aimed at preserving measures. By integrating the density-equalizing method with 3D quasi-conformal theory, we propose a parameterization approach that effectively balances both angular and authalic distortions.

\subsection{Proposed framework}
Let $\mathcal{M}$ be a 3D manifold in $\mathbb R^3$ with an outside boundary $\partial \mathcal{M}$. Our goal is to find a suitable parameterizing map: $f: \mathcal{M} \rightarrow \mathbb B \subset \mathbb R^3$ such that $f\left( \mathcal{M} \right)$ preserves certain properties. Here, $\mathbb B$ is the unit solid ball. 

To incorporate the quasi-conformal coefficients to the optimization problems, we define an auxiliary geometric coefficient $\overline{K}$ as follows
\begin{equation*}
    \overline{K} = \ln{K}, \quad K>0
\end{equation*}

For ease of computation, we divide the desired parameterization $f:\mathcal{M} \rightarrow \mathbb B$ into two parts, $f = \Bar{f} \circ g_0$, where $g_0: \mathcal{M} \rightarrow \mathbb B^0$ is the initial mapping of $\mathcal{M}$ onto the unit solid ball $\mathbb B^0$, and $\Bar{f}:\mathbb B^0 \rightarrow \mathbb B$ is a volumetric parameterization on the unit solid ball. Then, the problem of finding $f$ can be simplified as the problem of finding $\Bar{f}$ on the unit ball.

\textbf{Bijective solid ball quasi-conformal map}. We first consider the 3D quasi-conformal parameterizations. The energy formula for computing 3D quasi-conformal mappings is given by:
\begin{equation}\label{eqt:3DQC}
    \underset{f: \mathbb B^0 \rightarrow \mathbb B}{\arg \min } E_{\text{3DQC}}(f):=\underset{f: \mathbb B^0 \rightarrow \mathbb B}{\arg \min }\left\{\int_{\mathbb B^0} |\overline{K}(f) |^2\right\}
\end{equation}
subject to
\begin{align}
    \overline{K}\left(f\right)  \geq 0, \\
    \left(f \circ g_0 \right) \left( \partial \mathcal{M} \right) = \partial \mathbb B.
\end{align}
Here, $\overline{K}$ is the 3D geometric coefficient. $\partial \mathcal{M}$ and $\partial \mathbb B$ represent the outside boundary of the 3D manifold and unit ball, respectively.

Note that the above model focuses only on the geometric distortions. By minimizing the 3D geometry coefficient $\overline{K}$, Eq.~\eqref{eqt:3DQC} reduces the angle distortions.

\textbf{Bijective solid ball density-equalizing map}. Next, we consider measure-preserving parameterization. Based on the diffusion process, we propose an energy formula $E_{\text{3DDEM}}$ and obtain the optimal $\Bar{f}$ by minimizing it.
\begin{equation}\label{eqt:3DDEM}
    \underset{f: \mathbb B^0 \rightarrow \mathbb B}{\arg \min } \ E_{\text{3DDEM}}(f):=\underset{f: \mathbb B^0 \rightarrow \mathbb B}{\arg \min }\left\{\int_{\mathbb B^0} |\nabla \rho(f) |^2\right\}
\end{equation}
subject to
\begin{align}
    \overline{K}\left(f\right) \geq 0, \\
    \left(f \circ g_0 \right) \left( \partial \mathcal{M} \right) = \partial \mathbb B.
\end{align}
Here, $\rho$ denotes the density function and $\nabla \rho$ is the gradient of the density.

\textbf{Bijective solid ball density-equalizing quasi-conformal map}. In many practical situations, the parameterization result needs to consider not only the measure distortions but also the geometric errors. 
Inspired by this, we propose a variational model for finding the optimal $\Bar{f}$ that minimizes an energy functional $E_{\text{3DDEQ}}$ consisting of a density term and a geometric coefficient term, 
\begin{equation}\label{eqt:3DDEQ}
    \underset{f: \mathbb B^0 \rightarrow \mathbb B}{\arg \min } \ E_{\text{3DDEQ}}(f):=\underset{f: \mathbb B^0 \rightarrow \mathbb B}{\arg \min }\left\{\int_{\mathbb B^0} |\nabla \rho(f) |^2 + \alpha \int_{\mathbb B^0}| \overline{K}(f)|^2\right\}
\end{equation}
subject to
\begin{align}
    \overline{K}\left(f\right) \geq 0, \\
    \left(f \circ g_0 \right) \left( \partial \mathcal{M} \right) = \partial \mathbb B.
\end{align}
Here $\alpha$ is a user-defined positive weighting parameter for controlling the angle and density terms. 

Next, we prove the existence of the solution for the proposed models. 

For the bijective solid ball density-equalizing map, the existence of its solution is guaranteed by the property of the diffusion process. As $t \rightarrow \infty$, the diffusion flow will force the density $\rho$ to be fully equalized everywhere. At the same time, the gradient of the density $\nabla \rho$ will become $0$. Therefore, the solution of Eq.~\eqref{eqt:3DDEM} exists and can be obtained by solving the diffusion equation.

For the other two energy formulas, the existence of their solution is given by the following theorem.

\begin{theorem}\label{existence_sol}
    Suppose $\mathbb B$ is a unit solid ball. Let
    \begin{equation}
    \begin{aligned}
        \mathcal{A} = \{\overline{K} \in C^1(\mathbb B): & \ \|\overline{K}\|_{\infty} \leq c_1 ; \  \overline{K} \text{ is the 3D quasi-conformal coefficient} \\ & \text{of a 3D quasi-conformal map} \}
    \end{aligned}
    \end{equation}
    for some $c_1>0$. Then the models $E_{\text{3DQC}}$ and $E_{\text{3DDEQ}}$ admit a minimizer in $\mathcal{A}$. In fact, $\mathcal{A}$ is compact.
 \end{theorem}

\textit{Proof.}
To prove $\mathcal{A}$ is compact, we only need to prove that $\mathcal{A}$ is complete and totally bounded.

Note that $\overline{K} = 0$ is in the set $\mathcal{A}$, which indicates that $\mathcal{A}$ is nonempty. Now, we begin to prove the completeness of $\mathcal{A}$. Let $\{\overline{K}_n \}^{\infty}_{n=1}$ be a Cauchy sequence in $\mathcal{A}$ with respect to the norm $\| \|_{\infty}$. Since $\mathcal{A}\in L^{\infty}(\mathbb B)$ and $L^{\infty}(\mathbb B)$ is complete, there exists a $\Hat{K} \in L^{\infty}(\mathbb B)$, such that $\overline{K}_n \rightarrow K$ uniformly. Moreover, $\| \overline{K}_n \|_{\infty} < c_1 $ for each $n$ implies that $\| K \|_{\infty} < c_1 $ as well. Therefore, $ K \in \mathcal{A}$. Consequently, $\mathcal{A}$ is completely respectful to the norm $\| \cdot \|_{\infty}$.

To show $\mathcal{A}$ is totally bounded, we need to prove that for any $\epsilon>0$, there exists a finite collection of open balls with radius $\epsilon$, centered in $\mathcal{A}$, whose union covers $\mathcal{A}$. First, we construct a regular grid on $\mathcal{A}$ with edge length $\frac{1}{m}$. The re-indexing points of the net are denoted as $\{p_i \}_{i\in I}$. Then we define the smooth tent function as: $\mathcal{T}_{m,n,L} = \left\{ \left.\frac{k}{n}\Tilde{\delta}^{L}_{p_i} \right|_{\mathcal{D}^0}  \right\}_{i \in I,0 \leq k \leq n}$ such that: $\|\Tilde{\delta}^{L}_{p_i} - \delta^{L}_{p_i} \|_{\infty} <\frac{1}{L}$. Here, $\delta^{L}_{p_i}$ and $\Tilde{\delta}^{L}_{p_i}$ are the tent function and its smooth approximation, respectively. Now, consider the set $\mathcal{B}_{m,n,L} = \{\Tilde{K} \in \mathcal{A}: \Tilde{K} = \sum_i T_i, T_i \in \mathcal{T}_{m,n,L} \}$. Note that $\mathcal{B}_{m,n,L}$ contains finite elements. Therefore, for any given $\epsilon >0$, we can choose large enough $m,n$ and $L$, such that for any $\overline{K} \in \mathcal{A}$, there exists a tent function $\Hat{K} = \sum \frac{k}{n}\delta^{L}_{p_i}$ and $\Tilde{K} \in \mathcal{B}_{m,n,L}$ satisfying: $\| \overline{K} - \Hat{K} \|_{\infty} < \frac{\epsilon}{2}$ and $\| \Hat{K} - \Tilde{K} \|_{\infty} < \frac{\epsilon}{2}$. Then we can get: $\| \overline{K} - \Tilde{K} \|_{\infty} < \| \overline{K} - \Hat{K} \|_{\infty}  + \| \Hat{K} - \Tilde{K} \|_{\infty} < \epsilon$. Hence, $\mathcal{A} \subset \cup_{\Tilde{K} \in \mathcal{B}_{m,n,L}} B_{\epsilon}(\Tilde{K})$, where $B_{\epsilon}(\Tilde{K})$ is an open ball with center $\Tilde{K}$ and radius $\epsilon$. Consequently, $\mathcal{A}$ is totally bounded.

Since $\mathcal{A}$ is complete and totally bounded, $\mathcal{A}$ is compact. Therefore, $E_{\text{3DQC}}$ has a minimizer in $\mathcal{A}$. Moreover, note that $\nabla \rho$ is a $L^2$ function in $\mathbb B$ and $E_{\text{3DDEQ}}$ is continuous in $\mathcal{A}$. Hence, $E_{\text{3DDEQ}}$ has a minimizer in $\mathcal{A}$.
\hfill $\blacksquare$ 

With the theoretical minimizers of the models established, in Section~\ref{sec:main}, we propose several iterative schemes to obtain the final volumetric parameterizations $f:\mathcal{M} \rightarrow \mathbb B$ satisfying certain properties. While global minimizers exist theoretically, they may not fully adhere to the prescribed constraints in practice. Our iterative algorithms may therefore only converge to local minima.
\section{Proposed algorithms}\label{sec:main}
Let $\mathcal{M}$ be a 3D manifold represented as a tetrahedral mesh $\left(\mathcal{V},\mathcal{E},\mathcal{T}  \right)$, where $\mathcal{V}$
is the set of all vertices, $\mathcal{E}$ is the set of all edges, and $\mathcal{T}$ is the set of all tetrahedral faces. Our objective is to compute a bijective solid ball parameterization $f: \mathcal{M} \rightarrow \mathbb B$, guided by the prescribed energy formulas.

\subsection{Initial solid ball parameterization}\label{sec:initial_ball}
To begin, we first compute an initial solid ball map $\varphi:\mathcal{M} \rightarrow \mathbb B^0$ to parameterize the given 3D manifold $\mathcal{M}$ onto the unit ball $\mathbb B^0$. In this process, we employ harmonic maps to establish the parameterization, which is equivalent to solving the following system of equations:
\begin{equation}\label{eqt:Laplace}
    \left \{
    \begin{array}{ll}
    \Delta \varphi = 0     & \text{in} \quad \mathcal{M} \setminus \partial \mathcal{M}, \\
    \varphi \left(\partial \mathcal{M} \right) = \phi \left(\partial \mathcal{M} \right).    & 
    \end{array}    
    \right.
\end{equation}
This parameterization process can be simplified into two parts. First, we map the outside boundary onto the unit sphere to obtain the boundary conditions. Next, we construct the solid sphere based on these boundary conditions.

\subsubsection{Boundary condition}\label{ssec:boundary_map}
Let $\partial \mathcal{M}$ be the boundary of the given manifold $\mathcal{M}$. Note that $\partial \mathcal{M}$ is the genus-0 closed surface in $\mathbb R^3$. Our goal is to map $\partial \mathcal{M}$ onto the unit sphere $\mathbb S^2$. 

There are various methods available for computing the initial boundary mapping. Additionally, the outer boundary is crucial to the parameterization process. Consequently, it is essential to select appropriate outer boundaries for different energy formulations. In this context, we employ the SDEM method proposed by Lyu, Liu, and Choi \cite{lyu2024spherical} to compute the initial parameterization of the outer boundary.


SDEM is an iterative scheme specifically designed to achieve the bijective density-equalizing result for genus-0 closed surfaces. This process begins by conformally mapping the given surfaces onto a unit sphere using the Flash \cite{choi2015flash}. Subsequently, the method deforms the vertices by solving the diffusion equation, which can be interpreted as a diffusion flow. Furthermore, the quasi-conformal theory is induced to preserve the bijectivity during the iterative process. This method utilizes north-pole and south-pole stereographic projections and combines the map with a quasi-conformal correction to address overlaps. Finally, the updated map is projected back onto the sphere via the inverse stereographic projection. Readers are referred to \cite{lyu2024spherical} for more details.

\subsubsection{Interior map}\label{ssec:interior_harmonic}
Once the spherical boundary $\mathbb{S}^2$ is obtained, we impose it as a boundary constraint and compute an initial mapping $\phi \colon \mathcal{M} \to \mathbb{R}^3$ for the given manifold $\mathcal{M}$. Following the approach of \cite{wang2003volumetric}, we employ a harmonic map to fill the interior of the ball.

The harmonic map is a diffeomorphism that satisfies the Laplace equation:
\begin{equation}
    \Delta \varphi = 0.
\end{equation}
In theory, it's equivalent to minimizing the Dirichlet energy:
\begin{equation}
    E(\varphi) = \frac{1}{2} \int \|\nabla \varphi \|^2.
\end{equation}
Therefore, we only need to discretize the Dirichlet energy and find its minimum, which allows us to obtain the harmonic map on the tetrahedral mesh. 

\begin{figure}[t]
   \centering
   \includegraphics[width=\textwidth]{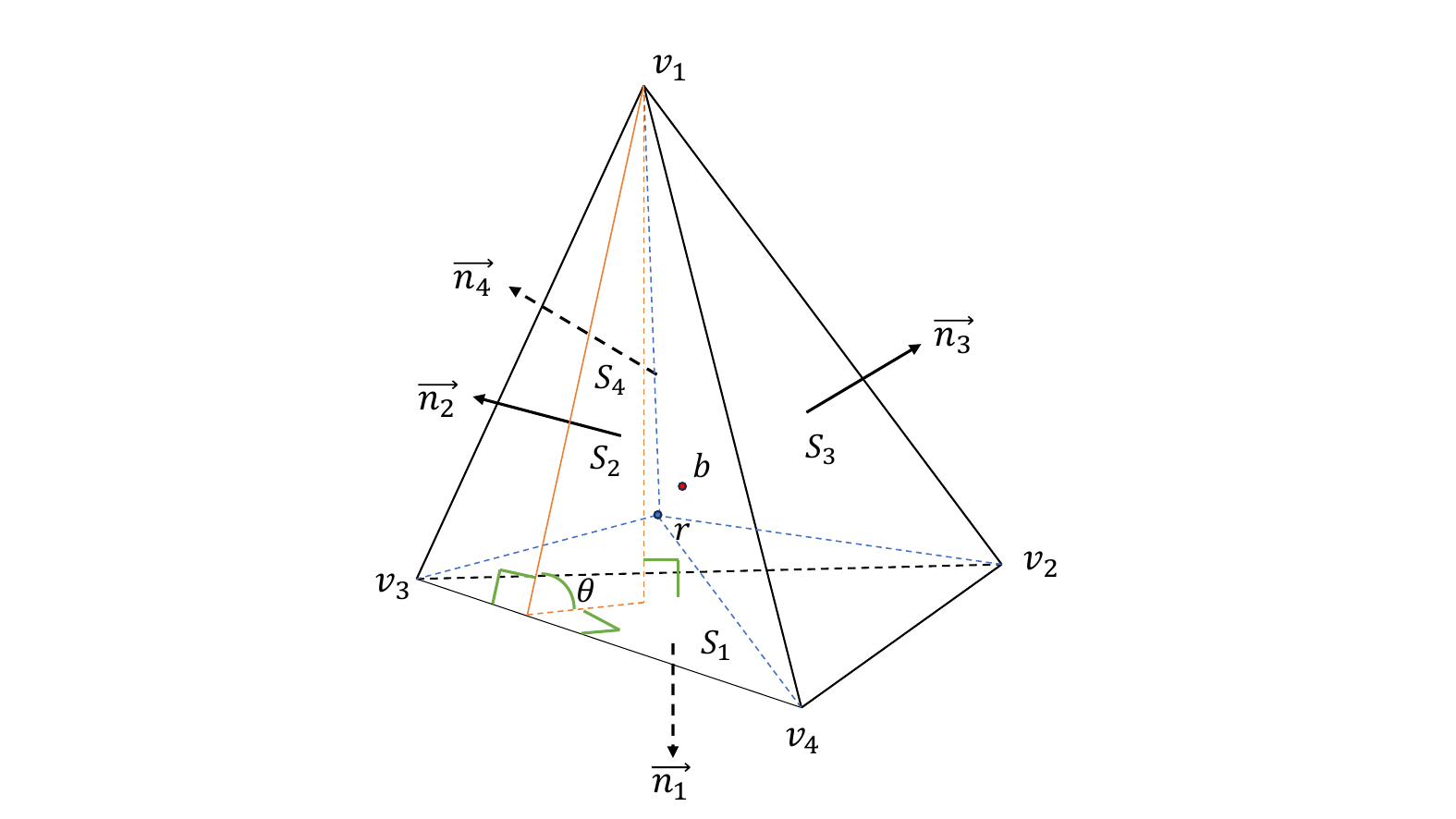}
    \caption{\textbf{An illustration of a tetrahedron.} }
    \label{fig:tet_element}
\end{figure}



Consider a tetrahedron $T$ with vertices $\{v_i\}_{i=1}^4$, faces $\{S_i\}_{i=1}^4$, and corresponding outward unit normal vectors $\{\vec{n}_i\}_{i=1}^4$, as shown in Fig.~\ref{fig:tet_element}. We first derive the expression for harmonic energy on this tetrahedral element.

Assume $\varphi$ is the given function on $T$, $b$ is the barycenter of $T$, and $r$ is the random point inside $T$. The barycentric coordinate of $\vec{r}$ is given by:
\begin{equation}
    \vec{r} = \sum^4_{i = 1}\left(\lambda_i \vec{v}_i \right).
\end{equation}
$\lambda_i$ is the barycentric parameter of $\vec{r}$ and can be computed by
\begin{equation}
\begin{aligned}
    \lambda_i & = \frac{V_i}{V} - \frac{1}{3V}\left<\vec{r}-\vec{b}, \vec{n}_i\right>\text{Area}\left(S_i \right)\\
    &=\frac{1}{4} - \frac{1}{3V}\left<\vec{r}-\vec{b}, \vec{S}_i\right>,
\end{aligned}
\end{equation}
where $V$ is the volume of $T$, $V_i$ is the corresponding volume of the vertex $v_i$, and, $\vec{S}_i = \text{Area}\left(S_i \right)\vec{n}_i$ . Therefore, the mapping result of $\vec{r}$ is 
\begin{equation}
    \varphi(\vec{r}) = \sum^{4}_{i = 1} \lambda_i \varphi(\vec{v}_i) = \sum^{4}_{i = 1}\left(\frac{1}{4} - \frac{1}{3V}\left<\vec{r}-\vec{b}, \vec{S}_i\right>\right)\varphi(\vec{v}_i).
\end{equation}
For simplicity, we use $\varphi_i$ to represent $\varphi(v_i)$ in the following.

The derivative can be obtained by
\begin{equation}
    \nabla \varphi = - \frac{1}{3V}\sum^{4}_{i = 1}\vec{S}_i \varphi(\vec{v}_i). 
\end{equation}

According to the property $\sum^{4}_{i = 1}\vec{S}_i = 0$ and $\left<\vec{S}_i,\vec{S}_j \right> = - \sum^{4}_{i \neq j}\left<\vec{S}_i,\vec{S}_j \right>, i = 1,2,3,4.$, the discrete harmonic energy on $T$ can be obtained by 
\begin{equation}
\begin{aligned}
E(\varphi) & =\frac{V}{2}<\nabla \varphi, \nabla \varphi>\\
&=\frac{1}{18 V}<\sum_{i=1}^4 \vec{S}_i \varphi_i, \sum_{i=1}^4 \vec{S}_i \varphi_i> \\
& =-\sum_{i \neq j} \frac{<\vec{S}_i, \vec{S}_j>}{18 V}\left(\varphi_i-\varphi_j\right)^2 \\
& =\sum_{\substack{i \neq j, p \neq q \\
\{p, q\}=I \backslash\{i, j\}}} l_{p q} \frac{\cot \left(\theta_{p q}\right)}{12}\left(\varphi_i-\varphi_j\right)^2 ,
\end{aligned}
\end{equation}
where $I = \{1,2,3,4 \}$. The last equation in the above holds due to
\begin{equation}
    -\frac{<\vec{S}_i, \vec{S}_j>}{V} = \frac{|\vec{S}_i|\frac{1}{2}l_{pq}h_q \cos{\theta_{pq}}}{\frac{1}{3}|\vec{S}_i|h_q\sin{\theta_{pq}}} = \frac{3}{2}l_{pq}\cot{\theta_{pq}},
\end{equation}
where $l_{pq}$ and $h_q$ are the edge length and height corresponding to the dihedral angle $\theta_{pq}$.

Then, to extend the formula into the tetrahedral mesh, we abbreviate the above formula as:
\begin{equation}\label{eqt:k_i_j}
    k_{u,v} = \sum_{i = 1}^{K} l_{i} \frac{\cot \left(\theta_{i}\right)}{12},
\end{equation}
where $\{ u,v\}$ is the edge shared by $K$ tetrahedral elements, $l_i$ is the length of the edge, and, $\theta_i$ is the corresponding dihedral angle. 
Therefore, the Laplacian operator on the tetrahedral mesh can be given by:
\begin{equation}\label{eqt:laplace_discrete}
    \Delta(\varphi) = \sum_{\{u,v \} }k_{u,v}\left(\varphi(v) - \varphi(u) \right).
\end{equation}

Based on the preceding discussion, the resulting mapping can be obtained by solving the following complex linear system:
\begin{equation}
    \left \{
    \begin{array}{ll}
    M^{H} \varphi(v) = 0     & \text{if} \quad v\in \mathcal{M} \setminus \partial \mathcal{M}, \\
    \varphi\left(\partial \mathcal{M} \right) = \phi \left( \partial \mathcal{M} \right).   & 
    \end{array}    
    \right.
\end{equation}
where 
\begin{equation}
    M^{H}_{ij} = \left \{
    \begin{array}{ll}
     k_{i,j}    & \text{if} \quad \left[v_i,v_j \right] \in \epsilon, \\
     -\sum_{l\neq j} k_{i,l}    & \text{if} \quad j = i, \\
     0      & \text{otherwise}. \\
    \end{array}
    \right.
\end{equation}
Here, the boundary mapping $\phi: \partial \mathcal{M} \rightarrow \partial \mathbb B^0$ is established through the SDEM method.


\subsubsection{Summary}
By combining the spherical boundary mapping from Section~\ref{ssec:boundary_map} with the interior harmonic mapping developed in Section~\ref{ssec:interior_harmonic}, we obtain our complete initial solid ball mapping algorithm. This algorithm effectively maps the input 3D manifold $\mathcal{M}$ into the unit ball $\mathbb{B}$, and its key steps are summarized in Algorithm~\ref{alg:Initial_ball}.

\begin{algorithm}[h]
\caption{Initial ball}
\label{alg:Initial_ball}
  \begin{algorithmic}[1]
    \Require A 3D manifold $\mathcal{M}$.

    \Ensure A solid ball map $\varphi:\mathcal{M}\to \mathbb{B}$.

\State Use SDEM to get the spherical outer boundary $\phi: \partial \mathcal{M} \rightarrow \partial \mathbb B$;

\State Compute the adjacency matrix $M^H$ such that 
$$   M^{H}_{ij} = \left \{
    \begin{array}{ll}
     k_{i,j}    & \text{if} \quad \left[v_i,v_j \right] \in \mathcal{E}, \\
     -\sum_{l\neq j} k_{i,l}    & \text{if} \quad j = i, \\
     0      & \text{otherwise}. \\
    \end{array}
    \right.;$$

\State Solve the linear system $    \left \{
    \begin{array}{ll}
    M^{H} \varphi(v) = 0     & \text{if}\quad v\in \mathcal{M} \setminus \partial \mathcal{M} \\
    \varphi\left(\partial \mathcal{M} \right) = \phi \left( \partial \mathcal{M} \right)   & 
    \end{array}    
    \right.$ and obtain the desired map $\varphi$;

  \end{algorithmic}
\end{algorithm}


\subsection{Bijective solid ball quasi-conformal map}\label{sec:3dqc}
After deriving the harmonic initial solid ball, our objective is to compute an optimal volumetric parameterization with lower geometric distortions. This is achieved minimizing the energy~\eqref{eqt:3DQC}, where $\overline{K} = \ln{K}$. Notably, reducing $\overline{K}$ corresponds to decreasing $K$. In the following, we proposed an iterative optimization scheme for solving the model~\eqref{eqt:3DQC}.

\subsubsection{Residual method}
Although gradient descent is a widely used optimization method, it is unsuitable for our problem. Specifically, the gradient descent direction of Eq.~\eqref{eqt:3DQC} is given by $-\overline{K}$, which only updates the ratio $\frac{a}{c}$, completely ignoring the middle term $b$. Since the eigenvalue $b$ also contains the geometric information, neglecting it during optimization may introduce undesirable distortions in the resulting parameterization.


We therefore propose a residual method that minimizes $K$ while accounting for all three eigenvalues ($a$, $b$, $c$), avoiding the geometric oversimplification of gradient descent. The details of the residual method are outlined as follows.

Let $\mathcal{T}^n = \{T^n_i\}$ be the $n$-th iteration's tetrahedral mesh with associated eigenvalues $\Lambda^n = \{(a^n_i, b^n_i, c^n_i)\}$, where $a^n_i \geq b^n_i \geq c^n_i$. As described in Eq.~\eqref{eqt:K}, the 3D quasi-conformal distortion of each tetrahedron is:
\begin{equation}
    K^n_i = \frac{a^n_i}{c^n_i}.
\end{equation}

To minimize geometric distortions, we first compute the residuals of the eigenvalues that quantify the current geometric deviation within the tetrahedral mesh. The residuals are defined as follows:
\begin{align}\label{eqt:residual}
    R^n_i = a^n_i - b^n_i, \\
    L^n_i = b^n_i - c^n_i,
\end{align}
where $R^n_i$ measures the deviation between the largest ($a^n_i$) and intermediate ($b^n_i$) eigenvalues, while $L^n_i$ captures the difference between the intermediate ($b^n_i$) and smallest ($c^n_i$) eigenvalues. These residuals provide an effective measure of the anisotropic distortion in each tetrahedral element.

When all residuals vanish (i.e., $R^n_i = L^n_i = 0$ for all $i$), the eigenvalues satisfy $a^n_i = b^n_i = c^n_i$ throughout the mesh. This condition implies that the 3D quasi-conformal coefficients $K^n_i$ are equal to 1 and the geometry coefficients $\overline{K}^n_i$ are equal to 0. Such a configuration represents an ideal, distortion-free parameterization of the mesh. Achieving this state is the fundamental objective of our optimization.

After obtaining $R^n_i$ and $L^n_i$, we utilize them to minimize the energy described in Eq.~\eqref{eqt:3DQC}. To address this problem, we first define the residual parameter:
\begin{equation}\label{eqt:res_para}
    \theta^n_i = \frac{K^n_i - 1}{ (K^n_i - 1) +\mathbf{C}}, 
\end{equation}
where $K^n_i$ is the 3D quasi-conformal coefficient and $\mathbf{C}$ is a constant that controls the residual weighting. 

The parameter $\theta$ plays an important role in the optimization process. Note that when $K = 1$, the residual parameter $\theta = 0$, implying that there is no distortion in the mapping. As the value of the 3D quasi-conformal coefficient $K^n_i$ increases, the residual parameter $\theta^n_i$ also increases accordingly. Moreover, the constant $C$ determines the sensitivity of $\theta^n_i$ to changes in $K^n_i$

By combining the residuals $\{R^n_i \}$, $\{L^n_i \}$ and the residual parameter $\{\theta^n \}$, we can update the $K^n$ according to the following formula:
\begin{equation}\label{eqt:update_3dqc_coeff}
    K^{n+1}_i = \frac{a^n_i - \theta^n_i R^n_i}{c^n_i + \theta^n_i L^n_i}.
\end{equation}
The corresponding eigenvalue updates are:
\begin{equation}
    \Lambda^{n+1}_i = \big(a^n_i - \theta^n_i R^n_i,\; b^n_i,\; c^n_i + \theta^n_i L^n_i\big).
\end{equation}
This residual-based approach provides an effective framework for minimizing geometric distortion while preserving the underlying geometric structure. Notably, according to the residual method, the larger the difference between two eigenvalues, the greater their respective changes after the update. This characteristic not only improves the accuracy of the method but also enhances its computational efficiency.

\subsubsection{Enforcing the bijectivity}
However, neither the initial mapping nor the residual method described above can guarantee bijectivity. In particular, when processing dense mesh regions, the solution to Eq.~\eqref{eqt:Laplace} may produce overlapping elements,

To address this issue, we propose a geometric correction scheme based on the distortion coefficient $K$ and eigenvalues $\Lambda$. As we mentioned in the previous section, $K<0$ if and only if the overlaps occur. Therefore, we check the geometric coefficient $K$ to find the overlaps. In order to correct the overlaps, we introduce a flipping function $\mathbb F$ for the geometric coefficient and eigenvalues:
\begin{equation}\label{eqt:flipping}
        \mathbb F(a,b,c) = \widetilde{\Lambda} = 
      \left \{
    \begin{array}{ll}
     (a,b,-c)    & \text{if} \quad K < 0 \\
     (a,b,c)     & \text{otherwise}. \\
    \end{array}  
    \right.
\end{equation}
Here, the $\widetilde{\Lambda}$ and $\widetilde{K}$ are the updated eigenvalues and geometric coefficient, respectively. 
Then, we can obtain a new mapping $\widetilde{g}$ by applying the following 3DQCS algorithm~\cite{chen2024novel3dmappingrepresentation}:
\begin{equation}\label{eqt:3dqcs_correct}
    \widetilde{g} = \textbf{3DQCS}(\widetilde{\Lambda}, \widetilde{K}).
\end{equation}
Here, $\widetilde{g}$ is a bijective solid ball parameterization with $\widetilde{K}_{\widetilde{g}} > 0$.

By incorporating this correction scheme at each iteration of the 3DQC mapping algorithm, we guarantee that the bijectivity condition is maintained throughout the optimization process. This ensures that all intermediate mappings remain foldover-free, while progressively reducing geometric distortion toward the optimal configuration.

\subsubsection{Summary}
With the computed 3D quasi-conformal coefficients, we update the mapping using the 3D quasi-conformal solver presented in~\cite{chen2024novel3dmappingrepresentation}. The stopping criterion is set as $E^{n+1}_{\text{3DQC}} - E^{n}_{\text{3DQC}} \leq 0$. The proposed algorithm is summarized in Algorithm~\ref{alg:3DQC}. In practice, the maximum number of iterations is $n_{\text{max}} = 100$ and the constant of residual parameter is $\mathbf{C} = 50$.

\begin{algorithm}[h]
\caption{3D quasi-conformal map}
\label{alg:3DQC}
  \begin{algorithmic}[1]
    \Require A 3D manifold $\mathcal{M}$, a residual constant $\mathbf{C}$.

    \Ensure A solid ball quasi-conformal map $f:\mathcal{M}\to \mathbb{B}$, where $\mathbb B$ is a solid ball..

\State Compute the initial solid ball parameterization $\varphi: \mathcal{M} \rightarrow \mathbb B^0$;

\State Compute the 3D quasi-conformal coefficients $\{K^0_i \}$ and the initial geometric distortions $\{\overline{K}^0_i \}$;

\State Set $n = 0$;

\Repeat

\State Compute the residuals $\{R^n_i \}$, $\{L^n_i \}$  by Eq.~\eqref{eqt:residual};

\State Compute the residual parameter $\{\theta^n_i \} $ by Eq.~\eqref{eqt:res_para};

\State Update the 3D quasi-conformal coefficient $\{K^{n+1}_i \}$ based on Eq.~\eqref{eqt:update_3dqc_coeff};

\State Update the geometric coefficient $\{\overline{K}^{n+1}_i \}$ by~\eqref{eqt:flipping};

\State Use the Eq.~\eqref{eqt:3dqcs_correct} to obtain a map $\widetilde{g}_{n+1}: \mathbb B^0 \rightarrow \mathbb B^{n+1} $;

\State Update n = n+1;

\Until $E^{n+1}_{\text{3DQC}} - E^{n}_{\text{3DQC}} \leq 0$ or \ $n \geq n_{\max}$;\\

\Return $f = \widetilde{g}_N \circ \varphi$, where $N$ is the total iteration number;
 
  \end{algorithmic}
\end{algorithm}

\subsection{Bijective solid ball density-equalizing map}\label{sec:3ddem}
In this section, building on Eq.~\eqref{eqt:3DDEM}, we propose a novel volumetric parameterization method for computing bijective density-equalizing maps in 3D cases. Let $\mathbb{B} \subset \mathbb{R}^3$ denote the unit solid ball with a positive population defined on it. The density $\rho$ is defined as the population per unit volume. In the following, we introduce our method for computing bijective density-equalizing maps utilizing diffusion flow and the 3D quasi-conformal theorem.

\subsubsection{Solid ball density-equalizing map}

In the discrete case, we use the finite element method to compute the volumetric density-equalizing map on the unit solid ball tetrahedral mesh. The density defined on each tetrahedral element can be computed by 
\begin{equation}
    \rho^\mathcal{T} = \frac{\text{Population}}{\text{Vol}(\mathcal{T})},
\end{equation}
where the $\text{Vol}(\mathcal{T})$ represents the volume of the corresponding tetrahedron. Note that our diffusion-based deformation occurs on the vertices. 
The vertex density defined on each vertex is given by:
\begin{equation}\label{eqt:conversion}
    \rho^\mathcal{V} = M^\mathcal{T}_{\mathcal{V}} \rho^\mathcal{T},
\end{equation}
where $M^\mathcal{T}_{\mathcal{V}}$ is a $|\mathcal{V}| \times |\mathcal{T}|$ tetrahedron-to-vertex matrix with:
\begin{equation}
\left(M^\mathcal{T}_{\mathcal{V}}\right)_{ij} = 
      \left \{
    \begin{array}{ll}
    \frac{\text{Vol}(\mathcal{T}_j)}{\sum_{\mathcal{T} \in \mathcal{N}^{\mathcal{T}}(\mathbf{x}_i)}\text{Vol}(\mathcal{T})}     & \text{if} \quad \mathcal{T}_i \quad \text{is incident to} \quad \mathbf{x}_i, \\
    0  & \text{otherwise}.
    \end{array}    
    \right.  
\end{equation}
Here $\mathcal{N}^{\mathcal{T}}(\mathbf{x}_i)$ denotes the neighboring tetrahedrons of the vertex $\mathbf{x}_i$. Alternatively, we can define the vertex-density $\rho^\mathcal{V}$ directly when working with point clouds, thus omitting Eq.~\eqref{eqt:conversion}. 

To solve Eq.~\eqref{eqt:duffusion}, we use the semidiscrete backward Euler method to discretize it into:
\begin{equation}\label{eqt:dis_diffusion}
    \frac{\rho^{\mathcal{V}}_{n+1} - \rho^{\mathcal{V}}_{n}}{\delta t} = \Delta_{n}\rho^{\mathcal{V}}_{n+1}.
\end{equation}
$\rho^{\mathcal{V}}_{n}$ is the $n$th iterative value of the vertex density $\rho^{\mathcal{V}}$, $\delta t$ is the time step for the operations, and, $\Delta_{n}$ is the FEM Laplace operator on the map. On the tetrahedral mesh, the Laplace operator $\Delta_{n}$ can be discrete as:
\begin{equation}
    \Delta_n = -A^{-1}_nL_n.
\end{equation}
Here, $A_n$ is a $|\mathcal{V}| \times |\mathcal{V}|$ diagonal matrix such that:
\begin{equation}\label{eqt:A_n}
    \left(A_n\right)_{ii} = \frac{1}{4} \sum_{\mathcal{T}_i \in \mathcal{N}^{\mathcal{T}}(\mathbf{x}_i)} \text{Vol}\left( \mathcal{T}_i \right)
\end{equation}
where $\mathcal{T}_i$ is a tetrahedral element in $\mathcal{N}^{\mathcal{T}}(\mathbf{x}_i)$ that contains the vertex $\mathbf{x}_i$. $L_n$ is the $|\mathcal{V}| \times |\mathcal{V}|$ mesh Laplacian matrix defined as:
\begin{equation}\label{eqt:L_n}
    \left(L_n\right)_{ij} = 
      \left \{
    \begin{array}{ll}
     k_{i,j}    & \text{if} \quad \left[\mathbf{x}_i,\mathbf{x}_j \right] \in \mathcal{E}, \\
     -\sum_{l\neq j} k_{i,l}    & \text{if} \quad j = i, \\
     0      & \text{otherwise}, \\
    \end{array}  
    \right.
\end{equation}
where $k_{i,j}$ is obtained by Eq.~\eqref{eqt:k_i_j} in Section~\ref{ssec:interior_harmonic}. Eq.~\eqref{eqt:dis_diffusion} is equivalent to:
\begin{equation}\label{eqt:diffusion_discrete}
    \rho^{\mathcal{V}}_{n+1} = \left(A_n + \delta t L_n \right)^{-1} \left( A_n \rho^{\mathcal{V}}_{n} \right).
\end{equation}
By solving the above equation, we can obtain the updated vertex density $\rho^{\mathcal{V}}_{n+1}$. To obtain the velocity $\mathbf{v} =-\frac{\nabla \rho^{\mathcal{V}}}{\rho^{\mathcal{V}}}$, we then need to compute the gradient $\nabla \rho^{\mathcal{V}}$ on every vertex to derive the velocity field. 

The discretization of the gradient $\nabla \rho$ on each tetrahedron is given by:
\begin{equation}
    \nabla \rho_n^{\mathcal{T}}(\mathcal{T}) = \frac{(\vec{e}_{kl}\times \vec{e}_{kj})\rho^{\mathcal{V}}_{n}\left(\mathbf{x}_i\right) + (\vec{e}_{ki}\times \vec{e}_{kl})\rho^{\mathcal{V}}_{n}\left(\mathbf{x}_j\right) + (\vec{e}_{ij}\times \vec{e}_{il})\rho^{\mathcal{V}}_{n}\left(\mathbf{x}_k\right) + (\vec{e}_{kj}\times \vec{e}_{ki})\rho^{\mathcal{V}}_{n}\left(\mathbf{x}_l\right)}{3\text{Vol}(\mathcal{T})},
\end{equation}
where $\mathcal{T} = [\mathbf{v_i},\mathbf{v_j},\mathbf{v_k},\mathbf{v_l}]$ is a tetrahedron and the edge cane be represented by $\vec{e}_{ij} = {[\mathbf{x}_i,\mathbf{x}_j]}$. Then, using Eq.~\eqref{eqt:conversion}, the $\nabla \rho_n^{\mathcal{T}}$ can be transferred into $\nabla \rho_n^{\mathcal{V}}$ by:
\begin{equation}
    \nabla \rho_n^{\mathcal{V}} = M^\mathcal{T}_{\mathcal{V}} \nabla \rho_n^{\mathcal{T}}.
\end{equation}
Therefore, the velocity field on every vertex can be computed by:
\begin{equation}\label{eqt:velocity_discrete}
    \mathbf{v}^{\mathcal{V}}_n = -\frac{\nabla \rho_n^{\mathcal{V}}}{\rho_n^{\mathcal{V}}}.
\end{equation}
As mentioned above, we aim to achieve the solid ball density-equalizing map while preserving the shape during iteration. In the continuous case, the $\mathbf{v}^{\text{bdy}}$ automatically lies on the tangent space of the unit sphere. However, to ensure the deformed boundary vertices are still located on the sphere, we employ the Riemann projection to the boundary velocity as follows:
\begin{equation}\label{eqt:velocity_projected_discrete}
    \widetilde{\mathbf{v}^{\mathcal{V}}_n}(\mathbf{x}_{\text{bdy}}) = \mathbf{v}^{\mathcal{V}}_n(\mathbf{x}_{\text{bdy}}) - \left( \mathbf{v}^{\mathcal{V}}_n(\mathbf{x}_{\text{bdy}}) \cdot \mathbf{n}^{\mathcal{V}}(\mathbf{x}_{\text{bdy}}) \right) \mathbf{n}^{\mathcal{V}}(\mathbf{x}_{\text{bdy}}),
\end{equation}
where $v_{\text{bdy}}$ is the boundary vertex, $\mathbf{n}^{\mathcal{V}}(v_{\text{bdy}})$ is the unit normal vector at the corresponding vertex. 

The descent direction of energy~\eqref{eqt:3DDEM} is given by:
\begin{equation}
    dE_{\text{3DDEM}} = 
      \left \{
    \begin{array}{ll}
        \widetilde{\mathbf{v}^{\mathcal{V}}_n}(\mathbf{x}_i)  & \text{if}\quad \mathbf{x}_i \in \partial \mathbb B^2 \\
    \mathbf{v}^{\mathcal{V}}_n(\mathbf{x}_i)     & \text{oterwise}, 
    \end{array}    
    \right.
\end{equation}
and the position of the vertices can be updated by:
\begin{equation}\label{eqt:update_map}
    f_{n+1} = f_n - \delta t \cdot dE_{\text{3DDEM}}.
\end{equation}
Although the Riemann projection on the boundary velocity can effectively remove the normal components, some vertices $\{\mathbf{x}_i\}$ may still move outside the unit spherical boundary following the deformation. To address this issue, we can normalize the boundary vertices $\mathbf{x}_{\text{bdy}}$ by their $L_2$ norm $\|\mathbf{x}_{\text{bdy}}\|$. As the iterations progress, the deformation of the vertices diminishes, indicating that the density approaches a constant value.

\subsubsection{Enforcing the bijectivity}\label{ssec:enfore_bijective}
As discussed in~\cite{lyu2024bijective,lyu2024spherical}, traditional density-equalizing maps cannot maintain bijectivity during the iteration process. Similar challenges arise in our solid ball parameterization framework, where the update rule in Eq.~\eqref{eqt:update_map} may generate tetrahedral overlaps under strong density gradients. Maintaining bijective mappings proves substantially more difficult for tetrahedral meshes compared to their triangular counterparts due to the increased topological complexity in three dimensions.

\begin{figure}[t]
   \centering
   \includegraphics[width=\textwidth]{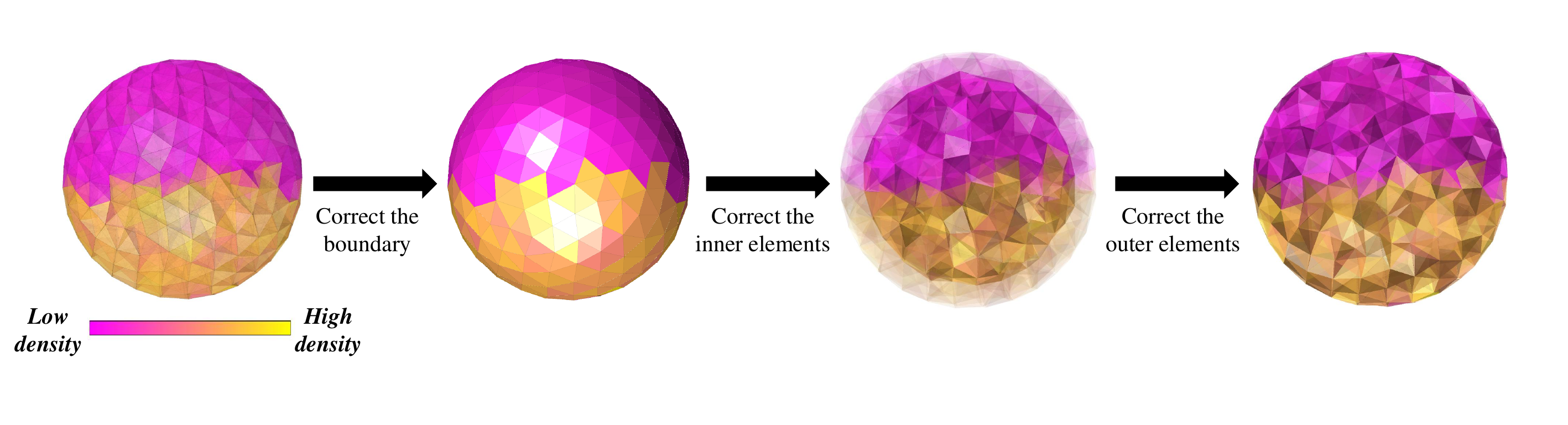}
    \caption{\textbf{An illustration of the correction scheme.} Given a solid ball, we first fix the overlaps on the outside triangle mesh using SDEM. Then, we correct the folding of the inner and outer tetrahedral elements by the 3D quasi-conformal theory, respectively.}
    \label{fig:correction_scheme}
\end{figure}

To address those issues, we proposed an overlap correction scheme (see Fig.~\ref{fig:correction_scheme}) grounded in 3D quasi-conformal theory. Our approach utilizes the 3D quasi-conformal coefficient $K$ to rectify the mesh folding caused by the deformation in Eq.~\eqref{eqt:update_map}. Let the initial ball be denoted as $\mathbb B^0$ and the ball resulting from the $n$th iteration be denoted as $\mathbb B^n$. Due to the complexity of tetrahedral meshes, our correction scheme is divided into several steps.

The correction process begins with the spherical boundary $\partial \mathbb{B}^n$, where density-gradient-driven vertex motion may induce overlaps.  To address those foldings, we employ the correction step in SDEM. Using SDEM's stereographic framework, we first project the mesh onto the complex plane, then compute and truncate the Beltrami coefficient $\mu$ to maintain $\|\mu\|_\infty < 1$ - the sufficient condition for bijectivity. The final inverse projection yields a valid spherical mesh, with full methodology detailed in~\cite{lyu2024spherical}.

With the bijective spherical boundary established, we now address foldings in the tetrahedral mesh $\mathbb{B}^n$. Analogous to the 2D case, tetrahedral foldovers correlate with negative 3D quasi-conformal coefficients $K^n_i$ - specifically, $K^n_i < 0$ indicates overlap in element $T^n_i$. 

To address the overlaps, we reverse the sign of the negative eigenvalue, resulting in a positive $K^n_i = \frac{a^n_i}{c^n_i}$. Here, $a^n_i,b^n_i$ and $c^n_i$ are resulting positive eigenvalues, subject to the condition $0\leq c^n_i \leq b^n_i \leq a^n_i$. However, this method fails to control the geometric distortions in the folded regions. We therefore develop an additional distortion minimization procedure, detailed in the following section.

To control the distortion of folded tetrahedra, we set a threshold $K_{T} > 1$ and enforce $K^n_i \leq K_T$ through an eigenvalue truncation operator $\mathbb{I}$:
\begin{equation}\label{eqt:truncation}
        \mathbb I(a,b,c) = 
      \left \{
    \begin{array}{ll}
     (a,b,c)    & \text{if} \quad K^n_i \leq K_{T} \\
     \left( \mathcal{G}_K(a), \mathcal{G}_K(b), \mathcal{G}_K(c) \right)     & \text{otherwise}, \\
    \end{array}  
    \right.
\end{equation}
where $\mathcal{G}_K$ is an affine transformation:
\begin{equation}
    \mathcal{G}_K(x) = l(x-e_\text{mean}) + e_\text{mean},
\end{equation}
where $e_\text{mean}$ is the mean value of the max and min eigenvalues:
\begin{equation}
    e_\text{mean} = \frac{a + c}{2}
\end{equation}
 and $l$ is the slope parameter:
\begin{equation}
    l = \frac{(K_T - 1)e_\text{mean}}{(K_T + 1)(a - e_\text{mean})}.
\end{equation}
Note that the threshold $K_{T} >1$ implies that $\mathcal{G}_K$ is monotonically increasing. Also, since
\begin{equation}
    l = \frac{(K_T - 1)e_\text{mean}}{(K_T + 1)(a - e_\text{mean})} < \frac{e_\text{mean}}{(a - e_\text{mean})} = \frac{-e_\text{mean}}{(c - e_\text{mean})}
\end{equation}
we have 
\begin{equation}
    \mathcal{G}_K(c) = l(c-e_\text{mean})e_\text{mean} > -e_\text{mean} + e_\text{mean} = 0.
\end{equation}
Therefore, we can obtain 
\begin{equation}
    \mathcal{G}_K(a) \geq \mathcal{G}_K(b) \geq \mathcal{G}_K(c) >0.
\end{equation}
 Moreover, it's clear that 
\begin{equation}
    \frac{\mathcal{G}_K(a)}{\mathcal{G}_K(c)} = \frac{\frac{2K_T e_\text{mean}}{K_T + 1}}{\frac{2e_\text{mean}}{K_T + 1}} = K_T.
\end{equation}

In conclusion, the correction step for the tetrahedral mesh $\mathbb B^n$ can be decoupled into two distinct parts: 
\begin{itemize}
    \item \textbf{Interior tetrahedrons}: Correct the overlaps in the interior tetrahedral elements in $\Hat{\mathbb{B}}^n$;
    \item \textbf{Boundary tetrahedrons}: Correct the overlaps on the boundary tetrahedral elements in $\Breve{\mathbb{B}}^n$.
\end{itemize}

For the interior mesh $\Hat{\mathbb{B}}^n$, we first identify all folded elements through their negative quasi-conformal coefficients $\{K^n_i\}_{i\in F}$, where $F$ denotes the index set of folded tetrahedrons in $\Hat{\mathbb{B}}^n$. Then, for each folded element $i\in F$, we reverse the sign of eigenvalues to obtain positive coefficients $\{\overline{K}^n_i\}_{i\in F}$. It is important to note that the values of $\{\overline{K}^n_i\}_{i\in F}$ may be large, indicating significant geometric distortions. Therefore, we finally apply the truncation function $\mathbb{I}$ to these $\overline{K}^n_i$ to reduce the geometric distortions.  

When $\overline{K}^n_i \leq K_T$, we retain the 3D quasi-conformal coefficient unchanged. Conversely, when $\overline{K}^n_i \geq K_T$, we substitute the eigenvalue $(\Bar{a}^n_i,\Bar{b}^n_i,\Bar{c}^n_i)$ into the affine translation $\mathcal{G}_K$ and obtain the $\left( \mathcal{G}_K(\Bar{a}^n_i), \mathcal{G}_K(\Bar{b}^n_i), \mathcal{G}_K(\Bar{c}^n_i) \right)$. The updated 3D quasi-conformal coefficient is $\overline{K}^n_i = \frac{\mathcal{G}_K(\Bar{a}^n_i)}{\mathcal{G}_K(\Bar{c}^n_i)} = K_T$. Here, for simplicity, we use $\overline{K}^n_i$ to represent the translation results as well.

Replacing the folding coefficient with the updated $\{\overline{K}^n_i\}_{i\in F}$, we obtained the new coefficients $\{\widetilde{K}^n_i \}_{i = 1}$, satisfying:
\begin{equation}
            \widetilde{K}^n_i = 
      \left \{
    \begin{array}{ll}
     K^n_i    & \text{if} \quad i \notin F, \\
     \overline{K}^n_i  &  \text{if} \quad i \in F. \\
    \end{array}  
    \right.
\end{equation}
Collaborated with the boundary $\partial \mathbb{B}^n$ computed by the above, we can obtain a new deformed map $\widetilde{g}_n$ by applying the 3DQCS algorithm~\cite{chen2024novel3dmappingrepresentation}:
\begin{equation}
    \widetilde{g}_n = \textbf{3DQCS}(\widetilde{\Lambda}^n, \widetilde{K}^n),
\end{equation}
with the outside boundary fixed. 

While the spherical boundary $\partial\mathbb{B}^n$ remains fixed during interior correction, this constraint introduces limitations for boundary-adjacent tetrahedrons in $\Breve{\mathbb{B}}^n$. Specifically, folding tetrahedrons that have either one face or one edge lying on the boundary may not be properly corrected in the above correction process. These geometric constraints can propagate numerical inaccuracies, leaving residual distortions in critical regions.


To address this issue, we fix the interior elements and repeat the aforementioned procedure to correct the mesh overlaps in $\Breve{\mathbb{B}}^n$. However, some small numerical errors may alter the shape of the boundary. To preserve the boundary's shape as a unit solid ball, we introduce an additional projection step for the boundary vertices. Ultimately, those steps lead to a bijective solid ball parameterization.

By incorporating this overlap correction scheme at each iteration of the solid ball density-equalizing mapping process, we can maintain bijectivity throughout the iterative process.

\subsubsection{Update the density}\label{ssec:redefine_density}
Our method deforms the solid ball through density-driven diffusion, moving vertices according to local density values and gradients each iteration. However, numerical errors can build up over time, and overlap corrections may shift vertices away from their ideal density-driven positions. To maintain accuracy, we add a final density update step each iteration to keep the deformation properly aligned with the density flow.


We denote the vertex position at $n$-th iteration as $\{\mathbf{x}^n_i \}$. Instead of directly updating the density obtained from~\eqref{eqt:diffusion_discrete}, we redefined the density on each tetrahedron as:
\begin{equation}\label{eqt:coupling}
    \rho^{\mathcal{T}}_{n+1}(T) = \frac{\text{Population}}{\text{Vol}\left([\mathbf{x}^{n+1}_i,\mathbf{x}^{n+1}_j,\mathbf{x}^{n+1}_k,\mathbf{x}^{n+1}_l ] \right)},
\end{equation}
where $T = [\mathbf{x}^{n+1}_i,\mathbf{x}^{n+1}_j,\mathbf{x}^{n+1}_k,\mathbf{x}^{n+1}_l ] \in \mathcal{T}$ is a tetrahedral element. After obtaining the new density $\rho^{\mathcal{T}}_{n+1}$, we utilize it in Eq.~\eqref{eqt:conversion} to update the vertex density and continue the iteration.

Instead of solving a single density diffusion equation over time, we recalculate a new equation at each iteration, updating the density and Laplace–Beltrami operator based on the current mapping. The density gradient is used only for that iteration, and an additional recoupling step improves the final mapping accuracy.

\subsubsection{Summary}
Combining together the initial solid ball parameterization (section~\ref{sec:initial_ball}) and the iterative scheme for solid ball density-equalizing mapping (Section~\ref{sec:3ddem}) with the overlap correction scheme (Section~\ref{ssec:enfore_bijective}) and the density recoupling scheme (Section~\ref{ssec:redefine_density}), we proposed the 3DDEM algorithm for computing solid ball density-equalizing mapping for 3D solid manifolds. Like~\cite{lyu2024spherical}, we set the stopping criterion as $\frac{\text{sd}(\rho_n^\mathcal{V})}{\text{mean}(\rho_n^\mathcal{V})} < \epsilon$, where $\epsilon$ is a stopping parameter. The algorithm is summarized in Algorithm~\ref{alg:3ddem}. In practice, the time step size is set to be $\delta t = 0.1$, the stopping parameter is set to be $\epsilon = 10^{-2}$, the maximum number of iterations is $n_{max} = 100$, and the threshold $K_T = 10$.

\begin{algorithm}[h]
\caption{Bijective solid ball density-equalizing map}
\label{alg:3ddem}
  \begin{algorithmic}[1]
    \Require A 3D solid manifold $\mathcal{M}$, a prescribed population, an initial solid ball parameterization $f_0:\mathcal{M} \to \mathbb{B}$, the stopping parameter $\epsilon$, the maximum number of iterations allowed $n_{\max}$, the threshold $K_T$, and residual parameter $\mathbf{C}$.

    \Ensure A solid ball density-equalizing map $f:\mathcal{M}\to \mathbb{B}$.

\State Compute the initial density $\rho^0_{\mathcal{T}}$ on $f_0(\mathcal{M})$ based on the prescribed population;

\State Set $n = 0$;

\State Let $\mathbf{x}_i(0) = f_0(v_i)$  for every vertex $v_i \in \mathcal{V}$;

\Repeat

\State Compute $A_n$ and $L_n$ using Eq.~\eqref{eqt:A_n} and Eq.~\eqref{eqt:L_n};

\State Obtain $\rho_{n+1}^{\mathcal{V}}$ by solving the diffusion equation~\eqref{eqt:diffusion_discrete};

\State Compute the velocity field ${\mathbf{v}}_{n+1}^{\mathcal{V}}$ using Eq.~\eqref{eqt:velocity_discrete};

\State Compute the projected velocity field $\widetilde{\mathbf{v}}_{n+1}^{\mathcal{V}}$ using Eq.~\eqref{eqt:velocity_projected_discrete};

\State Update the position of all $\mathbf{x}_i (t_{n+1})$ using Eq.~\eqref{eqt:update_map};




\State Apply the overlap correction scheme in Section~\ref{ssec:enfore_bijective} to further update $\mathbf{x}_i(t_{n+1})$;

\State Apply the re-coupling scheme in Section~\ref{ssec:redefine_density} to update $\rho_{n+1}^{\mathcal{T}}$ using Eq.~\eqref{eqt:coupling};

\State Update $n = n + 1$;

\Until $\frac{\text{sd}\left(\rho_{n}^{\mathcal{V}}\right)}{\text{mean}\left(\rho_{n}^{\mathcal{V}}\right)} < \epsilon$ \ or \ $n \geq n_{\max}$

\State The resulting spherical density-equalizing map is given by $f(v_i) = \mathbf{x}_i(t_n)$ for all $i$;

  \end{algorithmic}
\end{algorithm}

\subsection{Bijective solid ball density-equalizing quasi-conformal map}
After establishing the bijective 3D quasi-conformal mapping (3DQC) and bijective 3D density-equalizing map (3DDEM) frameworks, we now address the joint optimization of geometric and measure distortions under the unified model in Eq~\eqref{eqt:3DDEQ}. Our approach combines the strengths of the 3DQC and 3DDEM methods through an iterative scheme. For brevity, we denote $E_1 = \int \|\nabla \rho \|^2$, $E_2 = \alpha \int \|\overline{K} \|^2$, and derive the descent directions for each energy term to guide the optimization.

\subsubsection{Descent direction for 3DDEQ}
We begin by discussing duffusion term $E_1 = \int \|\nabla \rho \|^2$. In Section~\ref{sec:3ddem}, we propose an iterative method based on the diffusion theorem. This scheme works by iteratively adjusting the positions of vertices in the domain, resulting in a descent deformation that gradually minimizes the density variation and produces a density-equalizing outcome.

However, one major limitation of this vertex deformation method is that it is difficult to incorporate with the geometric information directly. This disconnect makes it difficult to achieve a unified framework that balances both density-equalization and geometric considerations. To address this issue, in this section, we reformulate the descent direction for the energy $E_1$ in terms of the 3D quasi-conformal coefficients $K$ and eigenvalues $\Lambda$.


Let $f$ be the result mapping with 3D quasi-conformal coefficient $K$  and eigenvalues $\Lambda = (a,b,c)$. As explained in Section~\ref{sec:3ddem}, the velocity induced by the density gradient is given by:
\begin{equation}
    df = - \frac{\nabla \rho^{\mathcal{V}}}{\rho^{\mathcal{V}}},
\end{equation}
where $\rho^{\mathcal{V}}$ is vertex density and $\nabla \rho^{\mathcal{V}}$ is the corresponding density gradient. Moreover, to maintain the shape as a unit solid ball, we remove the normal component $df^{\bot}_{\text{bdy}} = \left(df_{\text{bdy}}\cdot \mathbf{n}\right)\mathbf{n}$ of the boundary velocity $df$.

It is noteworthy that while $f$ experiences some perturbation, $K$ and $\Lambda$ will be adjusted concurrently. By substituting these adjustments into the 3D quasi-conformal representation formula~\cite{chen2024novel3dmappingrepresentation}, the adjustment can be expressed as follows:
\begin{equation}
    \mathcal{C}_{H}(\Lambda + d\Lambda_1) = f + df,
\end{equation}
where $\mathcal{C}_{H}$ denotes the 3D quasi-conformal solver associated with the eigenvalues $\Lambda$, and $H$ represents the source term. Moreover, $d\Lambda_1$ is the adjustment of the eigenvalues corresponding to the velocity $df$. By the inverse mapping theorem, the adjustment $d\Lambda_1$ can be obtained by:
\begin{equation}\label{eqt:eig_update_1}
    d\Lambda_1 = \mathcal{C}^{-1}_{H}(f + df) - \Lambda.
\end{equation}

Next, for the geometric term $E_2 = \alpha \int \|\overline{K} \| ^2$, we compute its descent direction using the residual method developed in Section~\ref{sec:3dqc}.Following an analogous approach to $E_1$, the descent direction $d\Lambda_2$ for $E_2$ can be computed by:
\begin{equation}\label{eqt:eig_update_2}
    d\Lambda_2 = \left(- \theta R, 0 , \theta L \right),
\end{equation}
where:
\begin{itemize}
\item $R$ and $L$ are the residuals defined in Eq.~\eqref{eqt:residual}
\item $\theta$ is the adaptive weighting parameter from Eq.~\eqref{eqt:res_para}.
\end{itemize}

By combining the descent directions derived for each energy component, we obtain the total descent direction for the combined energy $E_{\text{3DDEQ}} = E_1 + E_2$:
\begin{equation}
    d\Lambda = d\Lambda_1 + \alpha d\Lambda_2.
\end{equation}
where $d\Lambda_1$ is the descent direction derived from the diffusion term $E_1$,  $d\Lambda_2$ represents the descent direction derived from the geometric term $E_2$, and $\alpha > 0$ is the weighting parameter balancing between diffusion and geometric distortion control.

Using this combined descent direction, the updated eigenvalues can be computed as:
\begin{equation}
    \overline{\Lambda}^{n+1} = \Lambda^n + \delta t d\Lambda^n,
\end{equation}
where $\delta t$ is the time step size, $\Lambda^n$ eigenvalues of $n$th iterative map $f_n$, and $d\Lambda^n$ is the decent direction for $E_{\text{3DDEQ}}$ at $n$th itetation.

It is important to note that the bijectivity of the mapping cannot be guaranteed during the above steps. More specifically, some elements in $\overline{\Lambda}^{n+1}$ may be negative, inducing the overlaps in the mesh. 

To admit the issue, we employ the truncation function $\mathbb{I}$ in Eq.~\eqref{eqt:truncation} to the updated eigenvalues $\overline{\Lambda}^{n+1}$. The updated truncated eigenvalues are given by:
\begin{equation}\label{eqt:truncation_lambda}
    \Lambda^{n+1} = \mathbb{I} (\overline{\Lambda}^{n+1})=\left \{
    \begin{array}{ll}
     \Lambda_i^{n+1}    & \text{if} \quad K_i \leq K_{T}, \\
     \mathcal{G}_K(\Lambda_i^{n+1})     & \text{otherwise}, \\
    \end{array}  
    \right.
\end{equation}
where $\mathcal{G}_K(\Lambda_i^{n+1}) = \left(\mathcal{G}_K(a_i^{n+1}), \mathcal{G}_K(b_i^{n+1}), \mathcal{G}_K(c_i^{n+1}) \right)$ and $K_{T}$ is the threshold. The updated 3D quasi-conformal coefficients can be computed 
\begin{equation}\label{eqt:update_K}
    K^{n+1} = \frac{a^{n+1}}{c^{n+1}}.
\end{equation}
where $a^{n+1} = \max(\Lambda^{n+1})$ and $c^{n+1} = \min(\Lambda^{n+1})$. 

As aforementioned, the truncated result $\Lambda^{n+1}$ and $K^{n+1}$ guarantee the associated 3D quasi-conformal mapping to be bijective. This bijective modification map $f_{n+1}$ can be computed by:
\begin{equation}\label{eqt:midify_eq}
    f_{n+1} = \textbf{3DQCS}(\Lambda^{n+1}, K^{n+1}).
\end{equation}
Note that $\Lambda^{n+1}$ and $K^{n+1}$ integrate both measure-theoretic information and geometric constraints. Consequently, the resulting mapping $f_{n+1}$ balances both density and geometric distortions. By iteratively refining the mapping through updates to $K^{n+1}$ and $\Lambda^{n+1}$, we optimize the combined energy efficiently.

\subsubsection{Summary}
The detailed procedure of the iterative scheme is as follows. Suppose $f_n$ is the resulting map at the $n$th iteration with corresponding solid ball $\mathbb{B}^n$. The density $\rho_n^{\mathcal{V}}$, the 3D quasi-conformal coefficient $K^n$, and eigenvalues $\Lambda^n$  can be calculated by the same method as in the previous sections. We then update the $\Lambda^{n+1}$ and $K^{n+1}$, and compute the modification map $f_{n+1}$ as Eq.~\eqref{eqt:midify_eq}. We repeat the iterations until $|f_{n+1} - f_n|< \epsilon$, where $\epsilon$ is the stopping parameter.

The proposed algorithm is summarized in Algorithm~\ref{alg:3DDEQ}. In practice, the time step size is set to be $\delta t = 0.1$, the stopping parameter $\epsilon = 10^{-2}$, the maximum number of iterations is $n_{\text{max}} = 100$, the threshold $K_T = 10$, and the residual parameter $\mathbf{C} = 50 $.

\begin{algorithm}[h]
\caption{Bijective solid ball density-equalizing quasi-conformal map}
\label{alg:3DDEQ}
  \begin{algorithmic}[1]
    \Require A 3D solid manifold $\mathcal{M}$, a prescribed population, the stopping parameter $\epsilon$, the maximum number of iterations allowed $n_{\max}$, and the threshold $K_T = 10$.

    \Ensure A solid ball density-equalizing quasi-conformal map $f:\mathcal{M}\to \mathbb{B}$.
\State Compute the initial solid ball map $f_0: \mathcal{M} \rightarrow \mathbb{B}^0$, where $\mathbb{B}^0$ is a unit ball;

\State Compute the initial density $\rho_0^{\mathcal{T}}$, the initial 3D quasi-conformal coefficient $K^0$, and eigenvalues $\Lambda^0 = (a^0,b^0,c^0)$ on $f_0(\mathcal{M})$;

\State Set $n = 0$;

\Repeat

\State Apply the flow step (Eq.~\ref{eqt:update_map}) to get a modification map $\widetilde{f}_{n+1}$;

\State Compute the updated $\overline{\Lambda}^{n+1}$ by Eq.~\eqref{eqt:eig_update_1}--\eqref{eqt:eig_update_2};

\State Apply the truncation Eq.~\eqref{eqt:truncation_lambda} to obtain $\Lambda^{n+1}$;

\State Compute the updated $K^{n+1}$ by Eq.~\eqref{eqt:update_K};

\State Use the 3DQCR to reconstruct the $f_{n+1}:\mathbb{B}^0 \rightarrow \mathbb{B}^{n+1}$;

\State Update $\rho_{n+1}^{\mathcal{T}}$ based on $f_{n+1}$;

\State Update $n = n + 1$;

\Until $|f_{n+1} - f_n| < \epsilon$ \ or \ $n \geq n_{\max}$\\

\Return $f = f_N \circ \varphi$ where $N$ is the total number of iterations;

  \end{algorithmic}
\end{algorithm}

\section{Experiments}\label{sec:experiment}
In this section, we present experimental results to demonstrate the effectiveness of our 3DQC, 3DDEM, and 3DDEQ algorithms. The algorithms are implemented using MATLAB R2021a on the Windows platform. All experiments are conducted on a computer with an Intel(R) Core(TM) i9-13900T 1.10 GHz processor and 32 GB of memory. All manifolds are discretized in the form of tetrahedral meshes. In the following experiments, we set some slices on the tetrahedral mesh to show the results more clearly.

\subsection{Bijective solid ball quasi-conformal map}

\begin{figure}[t]
   \centering
   \includegraphics[width=0.8\textwidth]{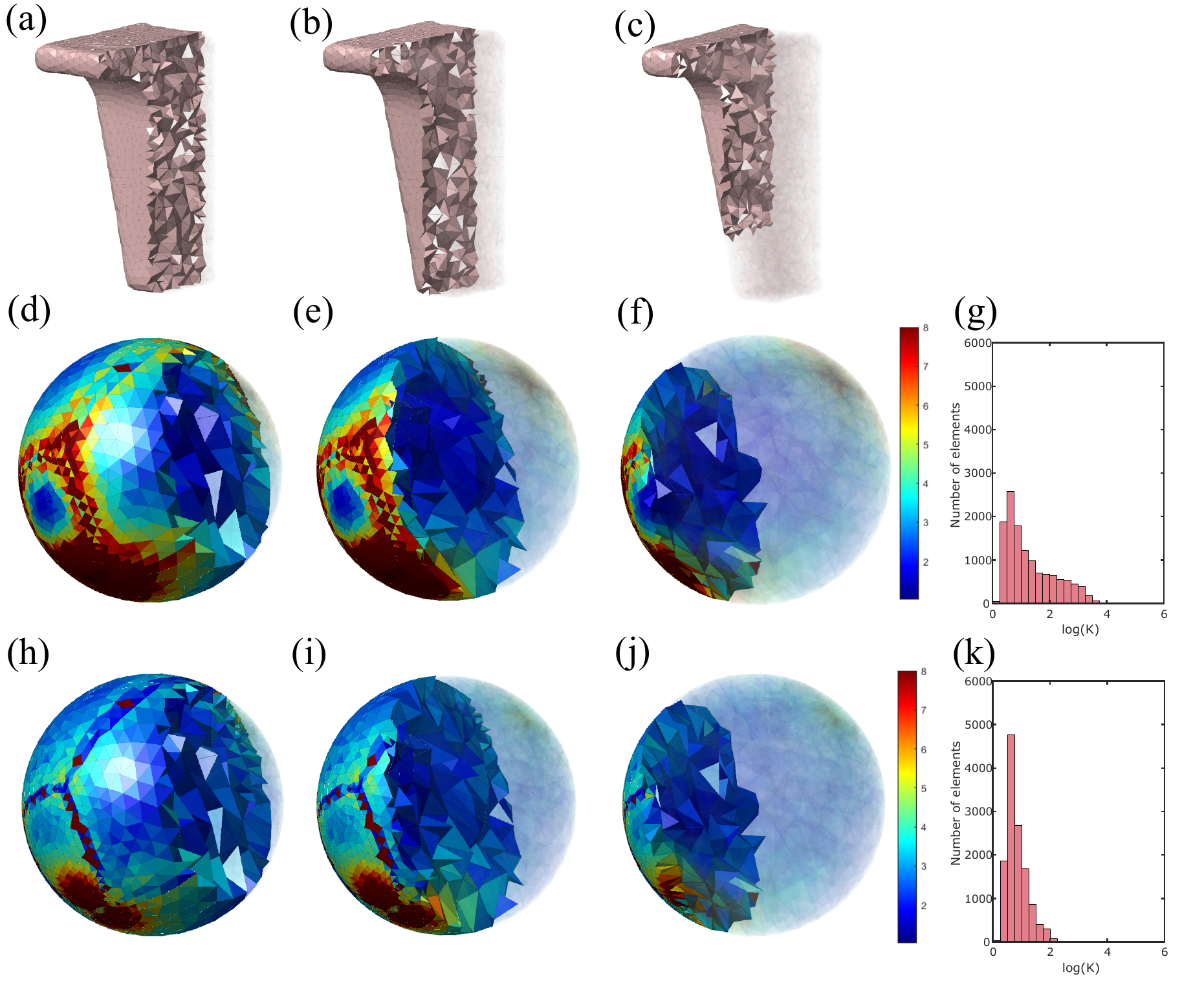}
    \caption{\textbf{3D quasi-conformal map on the T-Flex Dynamic module.} Top row (left to right): the original manifold with different slices. Middle row (left to right): the initial solid ball with different slices, and the histogram for initial geometric coefficients $\overline{K}$. Bottom row (left to right): the resulting solid ball with different slices, and the histogram for the resulting geometric coefficients $\overline{K}$.}
    \label{fig:conf_result1}
\end{figure}

We begin by testing our 3DQC algorithm on an industrial component: the T-Flex Dynamic module (Fig.~\ref{fig:conf_result1}(a)--(c)). The initial solid ball generated by Algorithm~\ref{alg:Initial_ball} is depicted in Fig.~\ref{fig:conf_result1}(d)--(f). In this visualization, the colormap is determined by the geometric coefficient $K$; specifically, areas of greater geometric distortion are represented with brighter colors.  Fig.~\ref{fig:conf_result1}(h)--(j) present the final results obtained using the proposed method. Notably, compared to the initial ball, the red regions with high geometric distortions in the final ball are reduced in size, while the blue regions with lower distortions are expanded. The histogram of the initial geometric coefficient in Fig.~\ref{fig:conf_result1}(g) indicates that the initial geometric distortion is significant. On the contrary, the final geometric coefficient in Fig.~\ref{fig:conf_result1}(k) is highly concentrated, suggesting that the final 3D quasi-conformal result efficiently preserves the geometric structure.

\begin{figure}[t]
   \centering
   \includegraphics[width=0.8\textwidth]{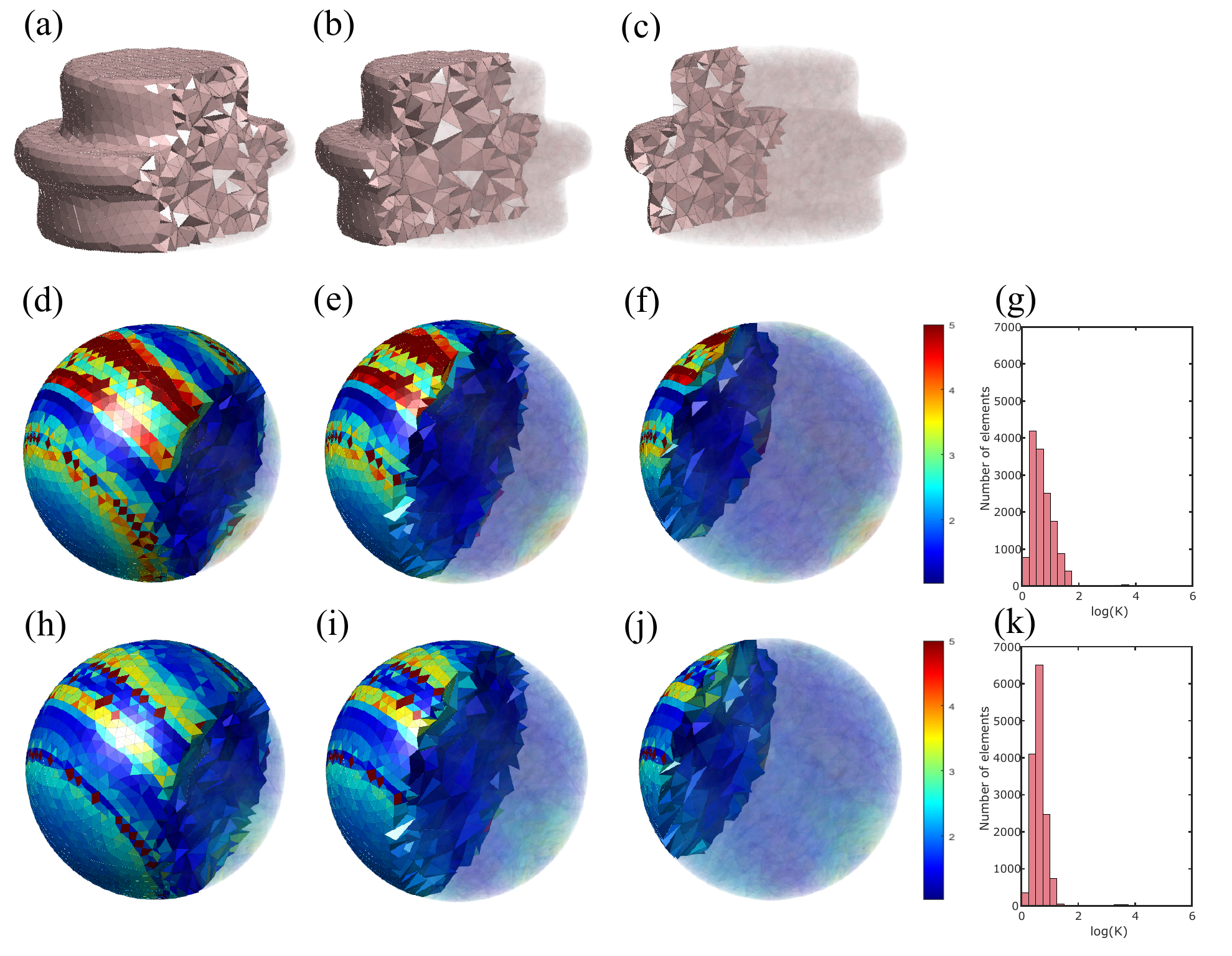}
    \caption{\textbf{3D quasi-conformal map on the Dual-Axis Guide Sleeve.} Top row (left to right): the original manifold with different slices. Middle row (left to right): the initial solid ball with different slices, and the histogram for initial geometric coefficients $\overline{K}$. Bottom row (left to right): the resulting solid ball with different slices, and the histogram for the resulting geometric coefficients $\overline{K}$.}
    \label{fig:conf_result2}
\end{figure}

Fig.~\ref{fig:conf_result2}(a)--(c) show another example of a Dual-Axis Guide Sleeve. It's easy to see that there are several high geometric distortion rings near the equator and the bottom in Fig.~\ref{fig:conf_result2}(d)--(f). After employing the 3DQC method, the result mapping is shown in Fig.~\ref {fig:conf_result2}(h)--(j), in which the high geometric distortion domain is shrunk significantly and the low distortion domain is enlarged. By considering the initial and final geometric coefficient histograms, we can see that the geometric distortions are efficiently reduced using our method.

\begin{figure}[t]
   \centering
   \includegraphics[width=0.8\textwidth]{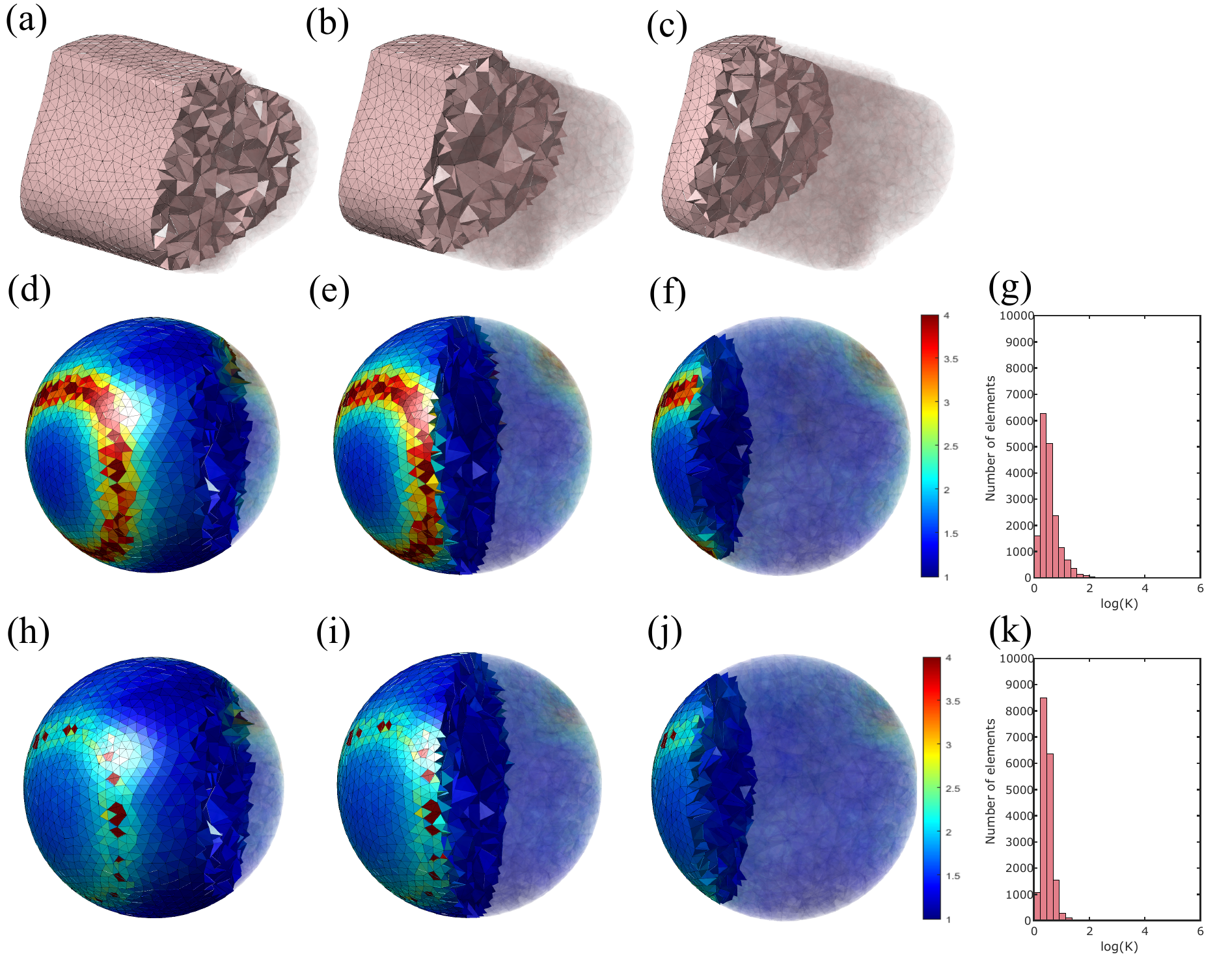}
    \caption{\textbf{3D quasi-conformal map on the engine model.} Top row (left to right): the original manifold with different slices. Middle row (left to right): the initial solid ball with different slices, and the histogram for initial geometric coefficients $\overline{K}$. Bottom row (left to right): the resulting solid ball with different slices, and the histogram for the resulting geometric coefficients $\overline{K}$.}
    \label{fig:conf_result3}
\end{figure}

The engine model is shown in Fig.~\ref{fig:conf_result3}(a)--(c). Fig.~\ref{fig:conf_result3}(d)--(f) present the initial ball colored by the 3D quasi-conformal coefficient $K$. From these figures, we can observe an annular domain with high geometric distortions on the left side. By applying the proposed 3DQC method to the initial solid ball, we obtain the result shown in Fig.~\ref{fig:conf_result3}(h)--(j). It is clear that the red domain is reduced, which means the geometric distortions are significantly reduced during the iteration. The initial and final geometric coefficient histograms are shown in Fig.~\ref{fig:conf_result3}(g) and (h). 

\begin{figure}[t]
   \centering
   \includegraphics[width=0.8\textwidth]{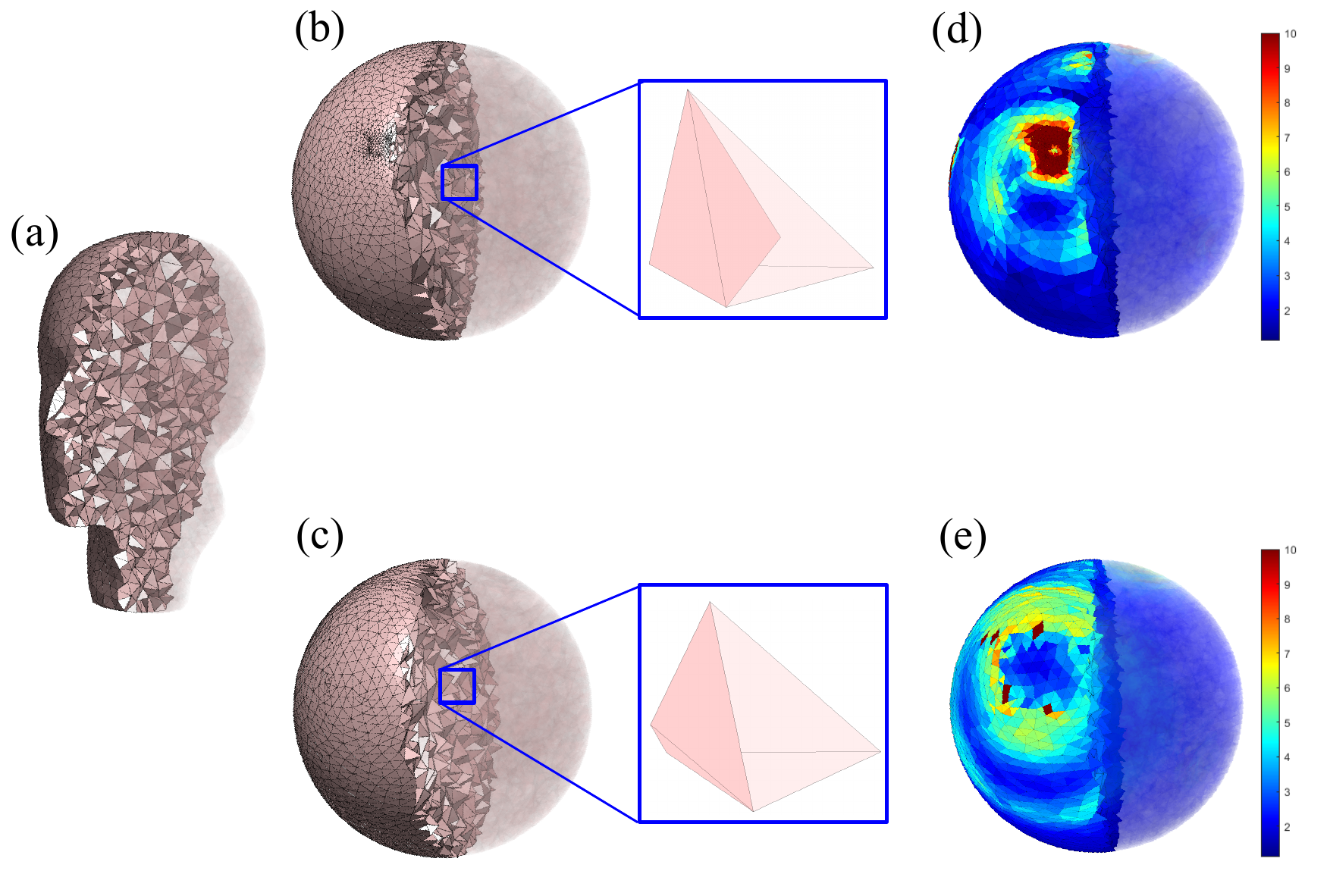}
    \caption{\textbf{Comparison between 3DQC and traditional harmonic method} (a) The Max Planck module. (b) The harmonic result with a zoom-in showing the overlap tetrahedron. (c) The 3DQC result with a zoom-in of the same tetrahedron. (d) The Harmonic result colored by the geometric coefficient $\overline{K}$. (d) The 3DQC result colored by the geometric coefficient $\overline{K}$. }
    \label{fig:conf_comparison}
\end{figure}

We further compare the performance of our 3DQC method and the existing harmonic method~\cite{gu2004genus} in Fig.~\ref{fig:conf_comparison}. Here, we consider mapping the Max Planck in Fig.~\ref{fig:conf_comparison}(a) using the two methods. It is evident from Figure~\ref{fig:conf_comparison}(b) that the harmonic method results in overlaps during the parameterization process. In contrast, the proposed 3DQC method maintains the bijectivity of the mapping, with no mesh overlaps, as shown in Figure~\ref{fig:conf_comparison}(c). The results colored by $K$ are presented in Fig.~\ref{fig:conf_comparison}(d) and (e). The result from the harmonic method exhibits several domains with high geometric distortions. In contrast, the geometric distortions in the 3DQC mapping are small due to the uniform distribution.

To provide a more quantitative analysis of our proposed method, we present detailed statistics in Table~\ref{tab:3DQC_comparison}. This table records the number of tetrahedral elements, as well as the means and standard deviations of the geometric coefficient $k$ obtained by the harmonic method and our 3DQC method, along with the number of overlaps. It can be observed that under our 3D quasi-conformal method,  both the mean values of the geometric coefficient and the standard deviation values are significantly reduced. Additionally, our method outperforms the harmonic approach in terms of the bijectivity of the mappings. This demonstrates that our method effectively reduces geometric distortions while preserving bijectivity.

\begin{table}[t]
  \caption{\textbf{Comparison between the traditional harmonic method and the 3DQC method for 3D manifolds.}}\label{tab:3DQC_comparison}
\begin{center}
\resizebox{\textwidth}{!}{
\begin{tabular}{|c|c|c|c|c|c|c|c|}
\hline
\multicolumn{1}{|c|}{ \multirow{2}*{\textbf{Manifold}} }& \multicolumn{1}{c|}{ \multirow{2}*{\textbf{\# tetrahedrons}}} & \multicolumn{3}{c|}{\textbf{Harmonic method}}&\multicolumn{3}{c|}{\textbf{3DQC}} \\
\cline{3-8}
\multicolumn{1}{|c|}{}&\multicolumn{1}{c|}{}  & \bf $\text{mean}(K)$ &\bf $\text{sd}(K)$ &\bf \# Overlaps & \bf $\text{mean}(K)$ &\bf $\text{sd}(K)$  & \bf \# Overlaps\\
\hline
L-cube & 9349 & 7.5657 & 59.1649 & 16 & 5.5085 & 22.3800 & 0    \\ \hline
Max Planck & 26639 &  3.5338 & 262.6341 & 12   & 2.2417 & 8.3775 & 0 \\ \hline
T-Flex Dynamic module & 14422  & 9.9609 & 93.9963 & 44  & 4.5769 & 8.4699 & 0  \\\hline
Engine Module  & 18003& 2.2887 & 10.1805 & 10  & 2.1199 & 8.9542 & 0 \\\hline
Dual-Axis Guide Sleeve  & 14422 & 2.7766 & 19.7058 & 13 & 2.2758 & 3.0060 & 0 \\\hline
\end{tabular}
}
\end{center}
\end{table}

    

\subsection{Bijective solid ball density-equalizing map}
\begin{figure}[t]
   \centering
   \includegraphics[width=0.7\textwidth]{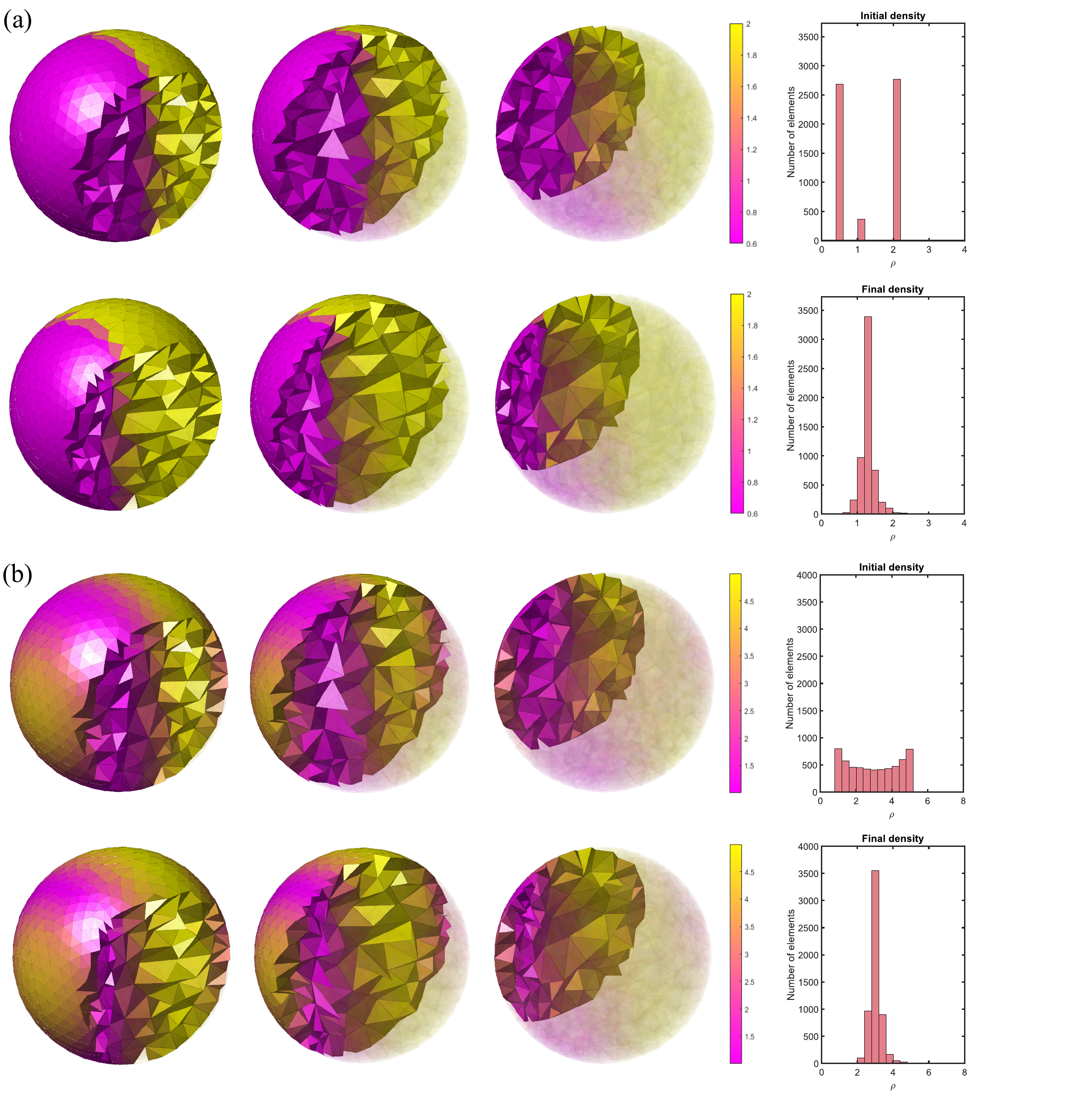}
    \caption{\textbf{3D density-equalizing map on solid balls.} (a) An example with discontinuous input density. (b) An example with continuous input density. For each example, the first row shows different slices of the initial map with the initial density histogram, and the last row shows different slices of the result map with the final density histogram. }
    \label{fig:dem_sythetic_result}
\end{figure}

Next, we test our 3DDEM algorithm for computing bijective density-equalizing mappings for solid 3D manifolds. We first consider some synthetic examples on unit solid balls. In Fig.~\ref{fig:dem_sythetic_result}(a), we define a discontinuous density distribution by assigning distinct population values to different regions of the ball. Specifically, the right semi-ball is initialized with a density four times greater than the left.  Additionally, in Fig.~\ref{fig:dem_sythetic_result}(b), we instead define a continuous density distribution across the solid ball, such that the maximum density is five times greater than the minimum density. We then apply the proposed method and obtain the solid ball density-equalizing mapping results shown in the figures. It is evident that the high-density regions in both examples are enlarged while the low-density regions are reduced. The initial and final density histograms demonstrate that our method effectively equalizes the density on the ball.

\begin{figure}[t]
   \centering
   \includegraphics[width=0.8\textwidth]{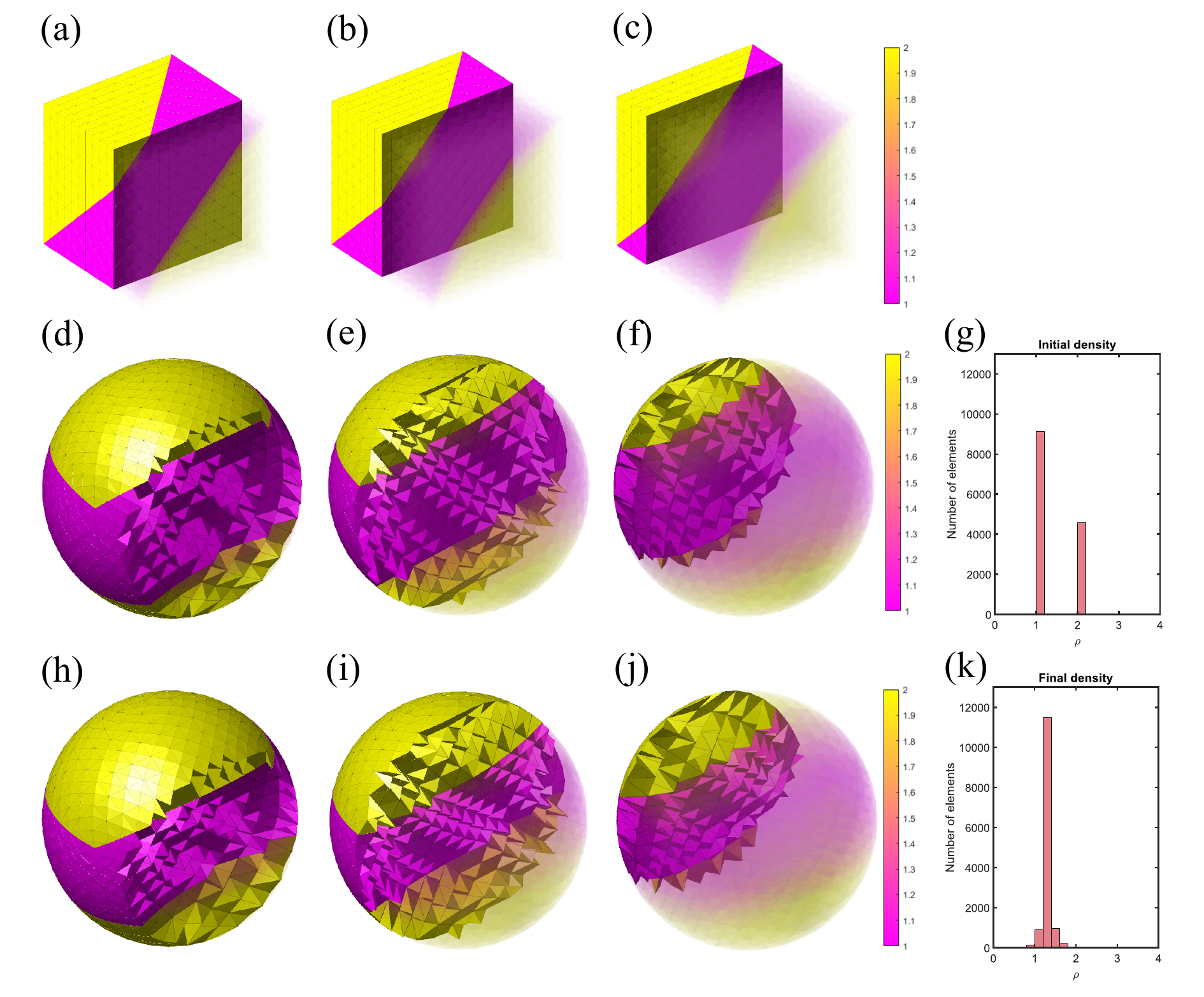}
    \caption{\textbf{3D density-equalizing map of cube.} Top row (left to right): the original manifold with different slices. Middle row (left to right): the initial solid ball with different slices, and the histogram for initial density $\rho_0$. Bottom row (left to right): the resulting solid ball with different slices, and the histogram for the final density $\rho$.}
    \label{fig:dem_cube}
\end{figure}

Then, we consider computing a 3D density-equalizing mapping of a cube. In Fig.~\ref{fig:dem_cube}(a)--(c), we use different slices to show the original cube. Here, we set different populations in different regions and apply our proposed algorithm to achieve a density-equalizing result. Fig.~\ref{fig:dem_cube}(d)--(f) show the initial solid ball obtained by the algorithm~\ref{alg:Initial_ball}. Fig.~\ref{fig:dem_cube}(h)--(j) show the final mapping result obtained by the 3DDEM algorithm. From those figures, we can see that the upper left and lower right domains are enlarged efficiently, while the interior domain is shrunk. Fig.~\ref{fig:dem_cube}(g) illustrates the density distortions of the initial solid ball, and the final density is shown in Fig.~\ref{fig:dem_cube}(k), which is highly concentrated at a constant value.

\begin{figure}[t]
   \centering
   \includegraphics[width=0.8\textwidth]{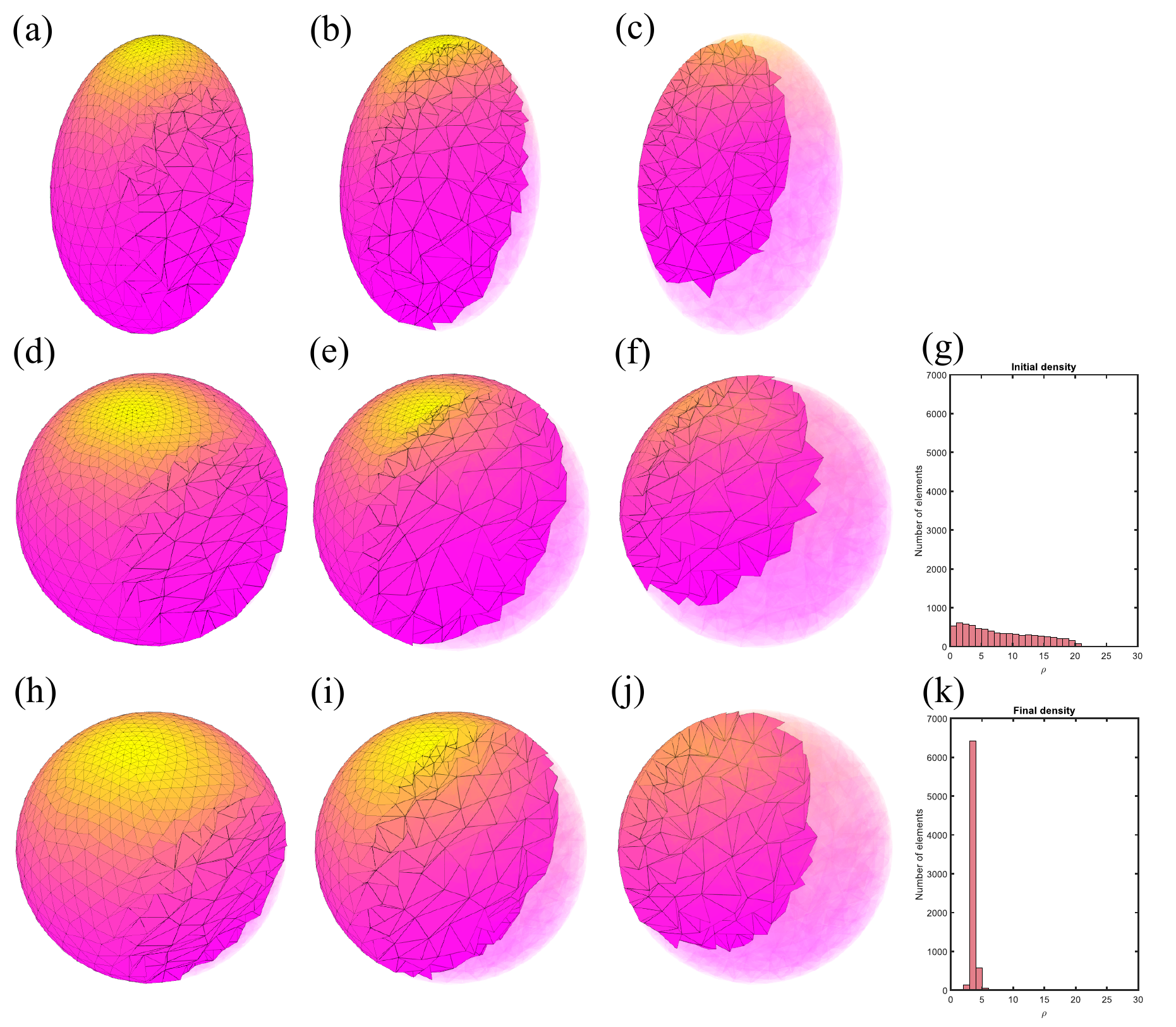}
    \caption{\textbf{3D density-equalizing map of Solid ellipsoid.} Top row (left to right): the original manifold with different slices. Middle row (left to right): the initial solid ball with different slices, and the histogram for initial density $\rho_0$. Bottom row (left to right): the resulting solid ball with different slices, and the histogram for the final density $\rho$.}
    \label{fig:dem_ellipse}
\end{figure}

It is natural to ask whether our proposed method can achieve  density-equalizing results on  non-uniform meshes. In Fig.~\ref{fig:dem_ellipse}(a)--(c), we consider a solid ellipsoid with a non-uniform tetrahedral mesh, where the mesh becomes increasingly sparse from top to bottom. We set the populations on the ellipsoid so that the density differs from the mesh distribution; specifically, the density increases as we move closer to the top. Fig.~\ref{fig:dem_ellipse}(d)--(f) show the different slices of the initial ball, from which we can see that the distribution of the mesh is similar to the original ellipsoid. Then we apply the proposed algorithm and obtain the final result shown in Fig.~\ref{fig:dem_ellipse}(h)--(j). Compared to the initial mapping, the final result enlarges the yellow domain and alleviates the concentration of elements in specific areas.  The histogram for the initial density in Fig.~\ref{fig:dem_ellipse}(h) indicates significant density distortions in the initial ball. In contrast, Fig.~\ref{fig:dem_ellipse}(k) shows that the final density is highly concentrated at a constant value, demonstrating that the final mapping effectively equalizes the density.

\begin{figure}[t]
   \centering
   \includegraphics[width=0.8\textwidth]{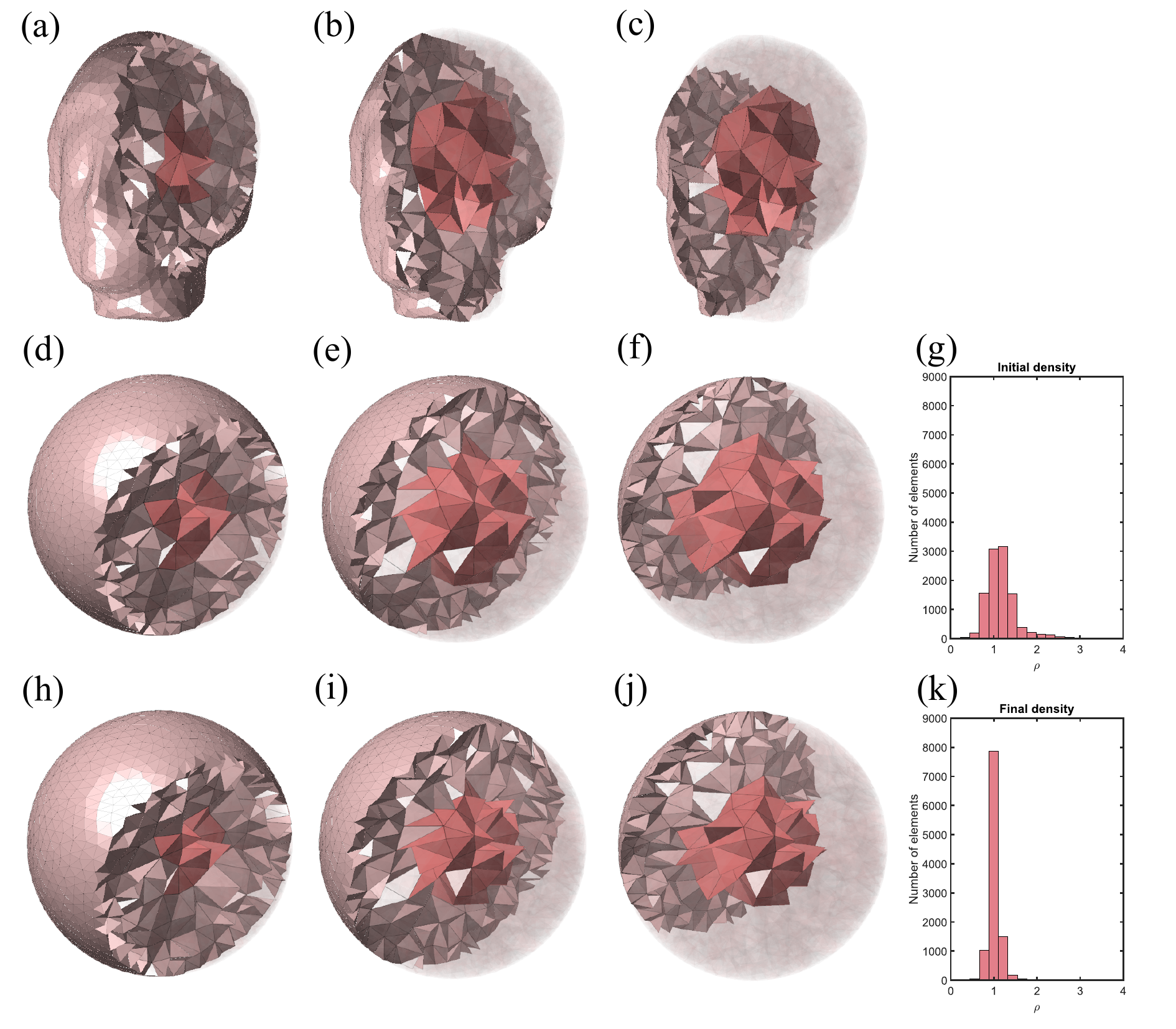}
    \caption{\textbf{3D density-equalizing map of Igea.} Top row (left to right): the original manifold with different slices. Middle row (left to right): the initial solid ball with different slices, and the histogram for initial density $\rho_0$. Bottom row (left to right): the resulting solid ball with different slices, and the histogram for the final density $\rho$.}
    \label{fig:dem_igea}
\end{figure}

Fig.~\ref{fig:dem_igea}(a)--(c) show an example for Igea. In this case, we set the population of the red region to be half the volume of each tetrahedral element in the original manifold. For the other regions, the population is set to match the volume of each tetrahedral element in the original manifold. The initial solid ball is depicted in Fig.~\ref{fig:dem_igea}(d)--(f). After applying the proposed method, the final result is presented in Fig.~\ref{fig:dem_igea}(h)--(j). Compared to the initial ball, it is clear that the red region has shrunk. The initial and final densities are shown in Fig.~\ref{fig:dem_igea}(g) and (h), indicating that the density has been effectively equalized.

\begin{figure}[t]
   \centering
   \includegraphics[width=0.8\textwidth]{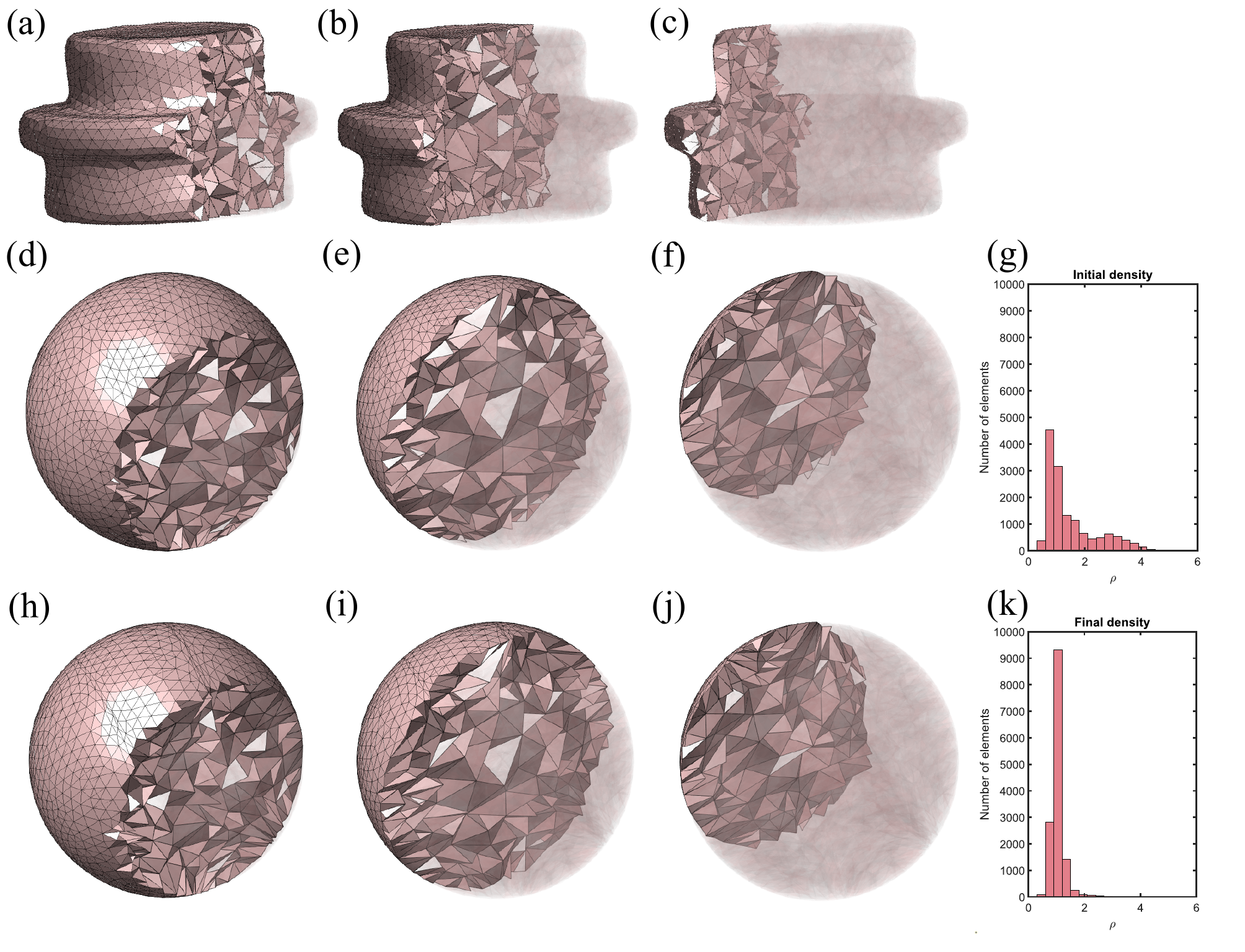}
    \caption{\textbf{3D volume-preserving parameterization of the Dual-Axis Guide Sleeve.} Top row (left to right): the original manifold with different slices. Middle row (left to right): the initial solid ball with different slices, and the histogram for initial density $\rho_0$. Bottom row (left to right): the resulting solid ball with different slices, and the histogram for the final density $\rho$.}
    \label{fig:dem_front}
\end{figure}

Note that by setting the population as the volume of each tetrahedron of the original manifold and applying our 3DDEM algorithm, we can obtain the solid ball volume-preserving parameterization. Fig.~\ref{fig:dem_front} shows the volume-preserving parameterization for the Dual-Axis Guide Sleeve. The original manifold is presented in Fig.~\ref{fig:dem_front}(a)--(c). Fig.~\ref{fig:dem_front}(d)--(f) are the slices of the initial ball. Using the 3DDEM method, the volume-preserving result is shown in Fig.~\ref{fig:dem_front}(h)--(j). From histograms~\ref{fig:dem_front}(g) and (k), we can see that the density is effectively equalized.  

For a more quantitative comparison, Table~\ref{tab:3DDEM} records the number of tetrahedral elements, variance of the initial density, variance of the final density, and number of overlaps for mapping different manifold models. It can be observed that after applying our 3DDEM method, the variance in density can be significantly reduced, demonstrating the efficiency and accuracy of our approach. Furthermore, all examples have zero overlaps, indicating that our method preserves bijectivity throughout the iteration process. 

\begin{table}[t!]
\small
    \caption{\textbf{The performance of our 3D density-equalizing method.} For each manifold, we record the number of tetrahedral elements, the variance of the normalized initial density $\Bar{\rho}_0 = \frac{\rho_0}{\text{mean}(\rho_0)}$, the variance of the normalized final density $\Bar{\rho} = \frac{\rho}{\text{mean}(\rho)}$, where $\rho_0$ is the initial vertex density and $\rho$ is the final vertex density and the number of overlaps.}\label{tab:3DDEM}
    
  \begin{center}
\resizebox{\textwidth}{!}{
  \begin{tabular}{|c|c|c|c|c|c|} \hline
    \bf Manifolds & \bf \# Elements  &\bf $\text{Var}(\Bar{\rho}_0)$  &\bf $\text{Var}(\Bar{\rho})$ &\bf \# Overlaps   \\\hline
    ball1  & 5828 & 0.2592 & 0.0003 & 0 \\ \hline
    ball2  & 5828 & 0.1948 & $<0.0001$ & 0  \\ \hline
    Cube & 13720 & 0.157 & 0.0016 & 0  \\ \hline
    Ellipsoid & 7250 & 0.5098 & 0.0014 & 0   \\ \hline
    Igea & 19539 & 0.1332 & 0.0003 & 0 \\ \hline
    Dual-Axis Guide Sleeve  & 14422 & 0.3562 & 0.0016 & 0 \\ \hline
  \end{tabular}
  }
\end{center}
\end{table}

\begin{figure}[t]
   \centering
   \includegraphics[width=0.6\textwidth]{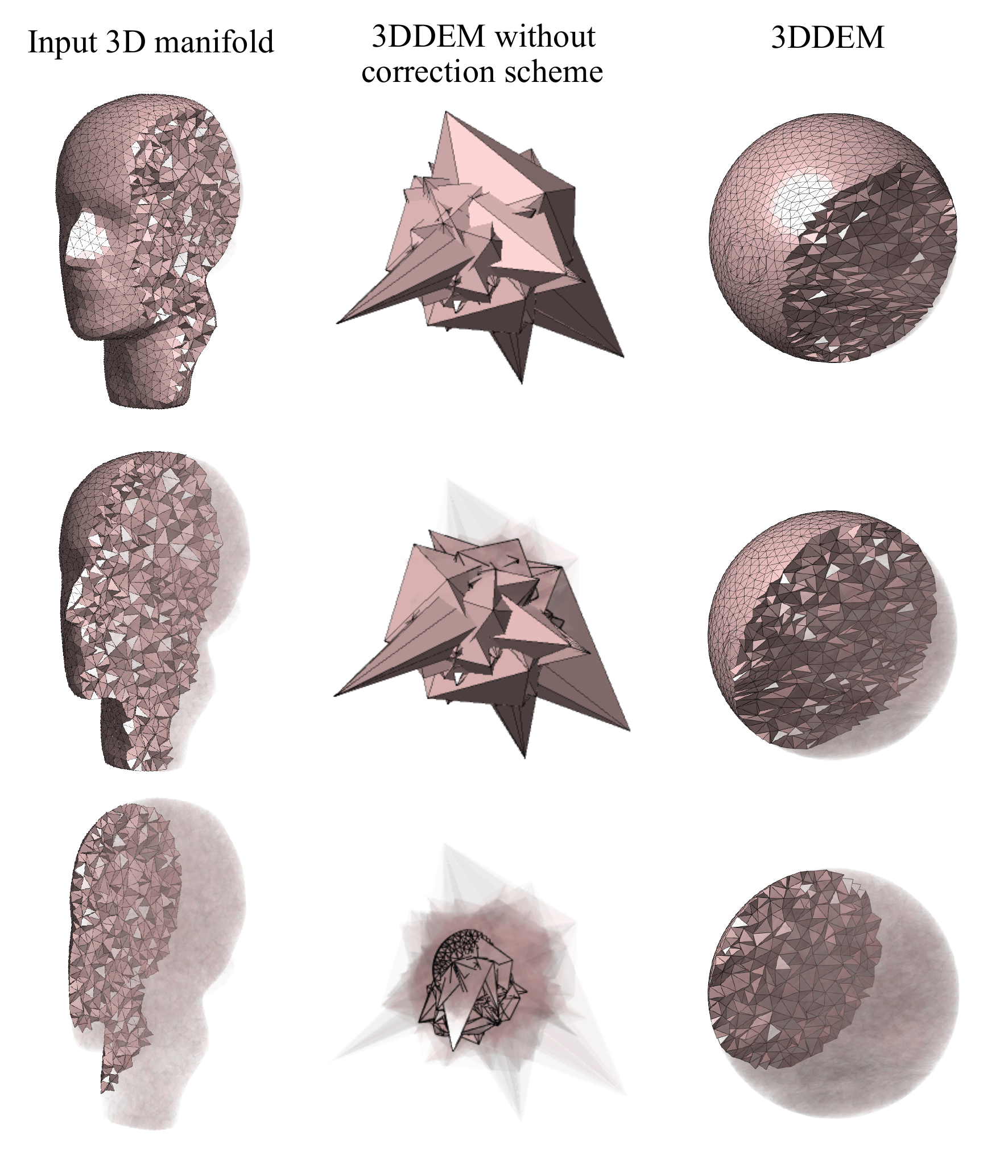}
    \caption{\textbf{Comparison between the proposed 3DDEM method and 3DDEM without the correction scheme.} Left column (top to bottom): the original manifold with different slices. Middle column (top to bottom): the solid density-equalizing map without the correction scheme. Right column (top to bottom): the solid density-equalizing map obtained by our proposed 3DDEM method.}
    \label{fig:MP_comparison}
\end{figure}

It is noteworthy that we have introduced a correction scheme in our 3DDEM method to enforce bijectivity and prevent overlaps during the iteration process. To highlight the importance of this correction scheme, we compare the mapping results obtained by our 3DDEM method with those from 3DDEM without the correction scheme in Fig.~\ref{fig:MP_comparison}. Without the correction scheme, the resulting mapping exhibits significant folding and fails to preserve the spherical shape. Additionally, due to overlaps, this result is not density-equalizing at all. In contrast, the mapping result obtained by our proposed 3DDEM method with the correction scheme shows no folding. This comparison demonstrates the critical role of the correction scheme in maintaining bijectivity and enhancing the efficiency of the mapping process.

\subsection{Bijective solid ball density-equalizing quasi-conformal map}
\begin{figure}[t]
   \centering
   \includegraphics[width=0.8\textwidth]{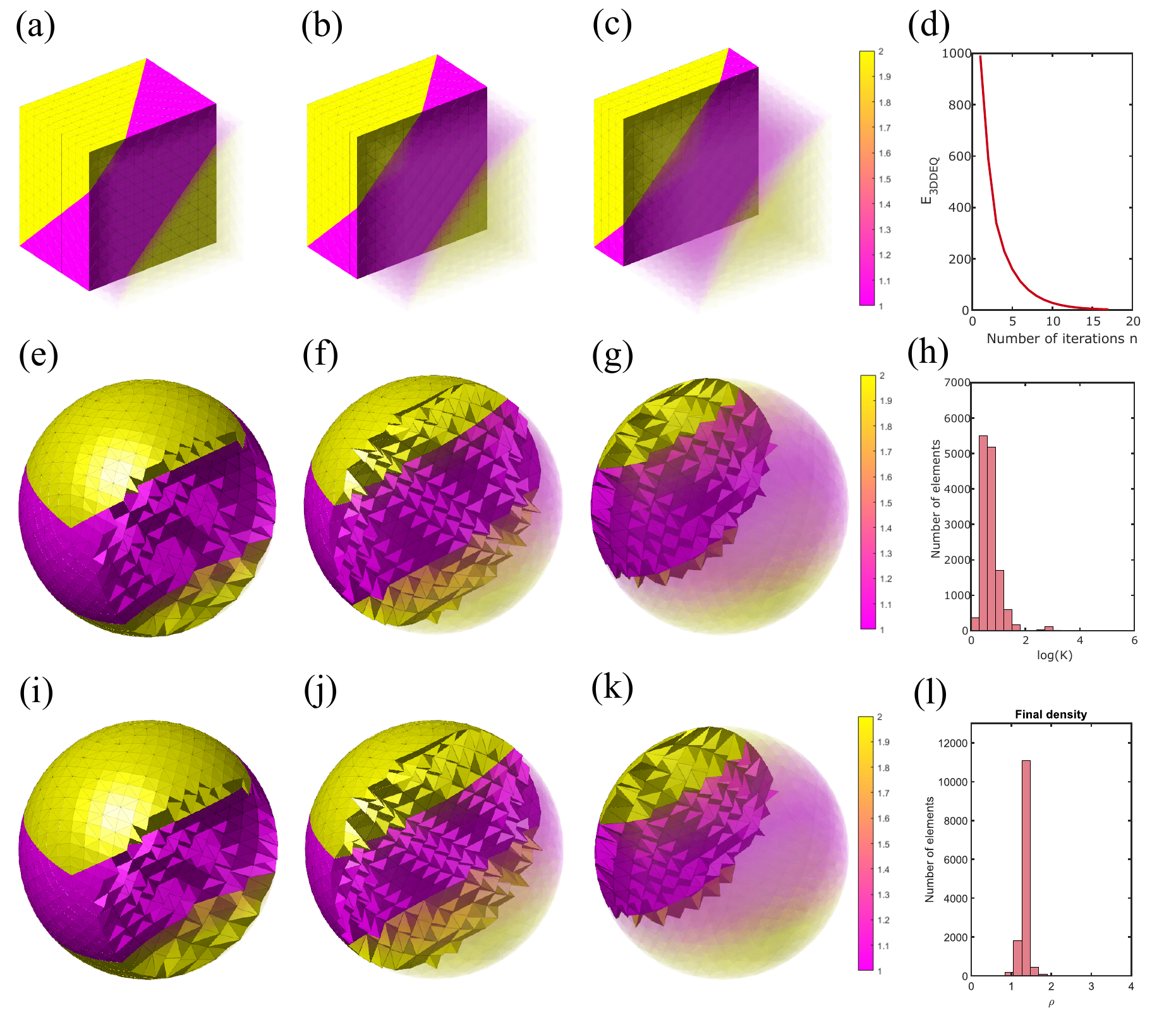}
    \caption{\textbf{3D density-equalizing quasiconformal map of cube.} Top row (left to right): the original manifold with different slices, and the histogram of the energy $E_{\text{3DDEQ}}$. Middle row (left to right): the initial solid ball with different slices, and the histogram for the geometric coefficient $\overline{K}$. Bottom row (left to right): the result solid ball with different slices, and the histogram for the final density $\rho$.}
    \label{fig:deq_cube}
\end{figure}

In this part, we test our 3DDEQ algorithm for computing 3D density-equalizing quasiconformal maps. In all experiments, we set $\alpha = 0.01$. We first consider a solid cube with the input populations shown in Fig.~\ref{fig:deq_cube}(a)--(c). Note that the population is identical to the one used in the example in Fig.~\ref{fig:dem_cube}. The initial and final results are presented in Fig.~\ref{fig:deq_cube}(e)--(g) and (i)--(k). Fig.~\ref{fig:deq_cube}(d) illustrates the change in energy $E_{\text{3DDEQ}}$ throughout the iterations, demonstrating that our method converges rapidly. The histogram of geometric coefficient $\overline{K}$ is shown in Fig.~\ref{fig:deq_cube}(h), from which we can observe that the geometric distortions of result mapping are small. Moreover, the final density in Fig.~\ref{fig:deq_cube}(l) highly concentrates to a constant, indicating that our proposed method effectively equalizes the density.

\begin{figure}[t]
   \centering
   \includegraphics[width=0.8\textwidth]{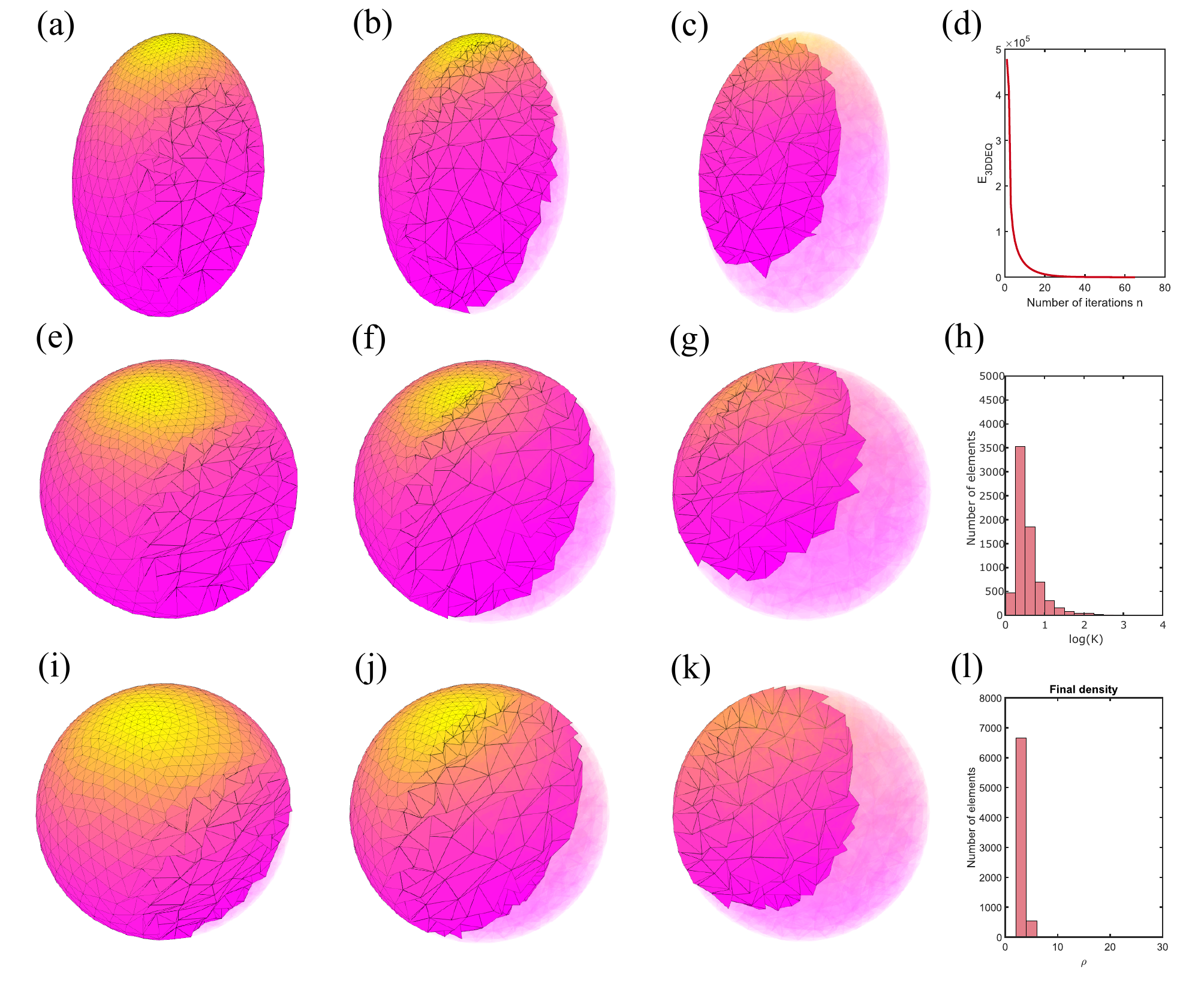}
    \caption{\textbf{3D density-equalizing quasiconformal map of solid ellipsoid.} Top row (left to right): the original manifold with different slices, and the histogram of the energy $E_{\text{3DDEQ}}$. Middle row (left to right): the initial solid ball with different slices, and the histogram for the geometric coefficient $\overline{K}$. Bottom row (left to right): the result solid ball with different slices, and the histogram for the final density $\rho$.}
    \label{fig:deq_ellipse}
\end{figure}

Fig.~\ref{fig:deq_ellipse} shows an example of mapping a non-uniform solid ellipsoid using our 3DDEQ method. Here, the input population is the same as in the last section. From the results, we can observe that our 3DDEQ method can handle the non-uniform example very well. The energy histogram in Fig.~\ref{fig:deq_ellipse}(d) shows that the proposed method can reduce the energy efficiently. The final geometric coefficient and density in Fig.~\ref{fig:deq_ellipse}(h) and (l) reflect that the 3DDEQ algorithm can both reduce the geometric and density distortions significantly.

\begin{figure}[t]
   \centering
   \includegraphics[width=0.8\textwidth]{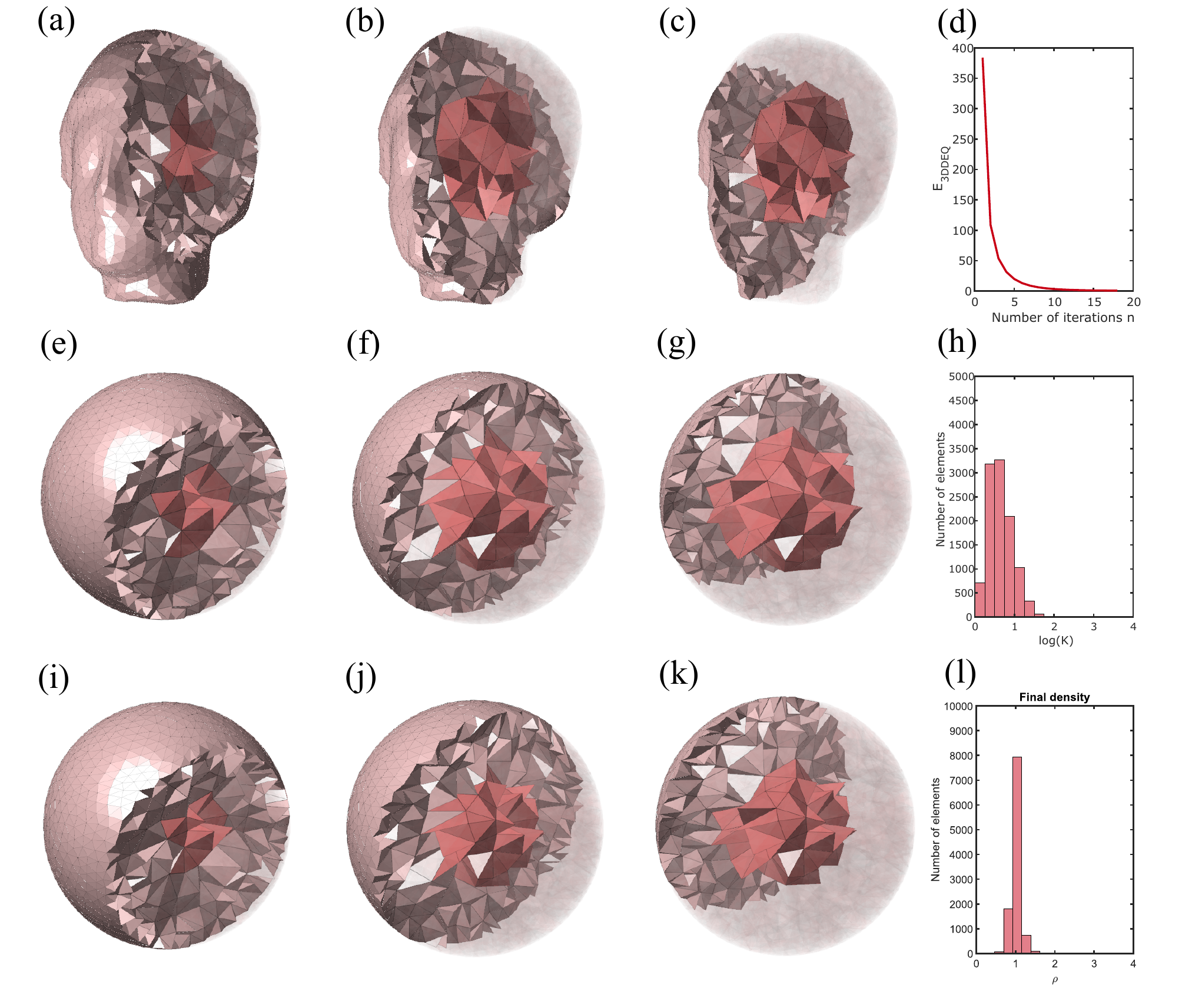}
    \caption{\textbf{3D density-equalizing quasiconformal map of Igea.} Top row (left to right): the original manifold with different slices, and the histogram of the energy $E_{\text{3DDEQ}}$. Middle row (left to right): the initial solid ball with different slices, and the histogram for the geometric coefficient $\overline{K}$. Bottom row (left to right): the result solid ball with different slices, and the histogram for the final density $\rho$.}
    \label{fig:deq_igea}
\end{figure}

\begin{figure}[t]
   \centering
   \includegraphics[width=0.8\textwidth]{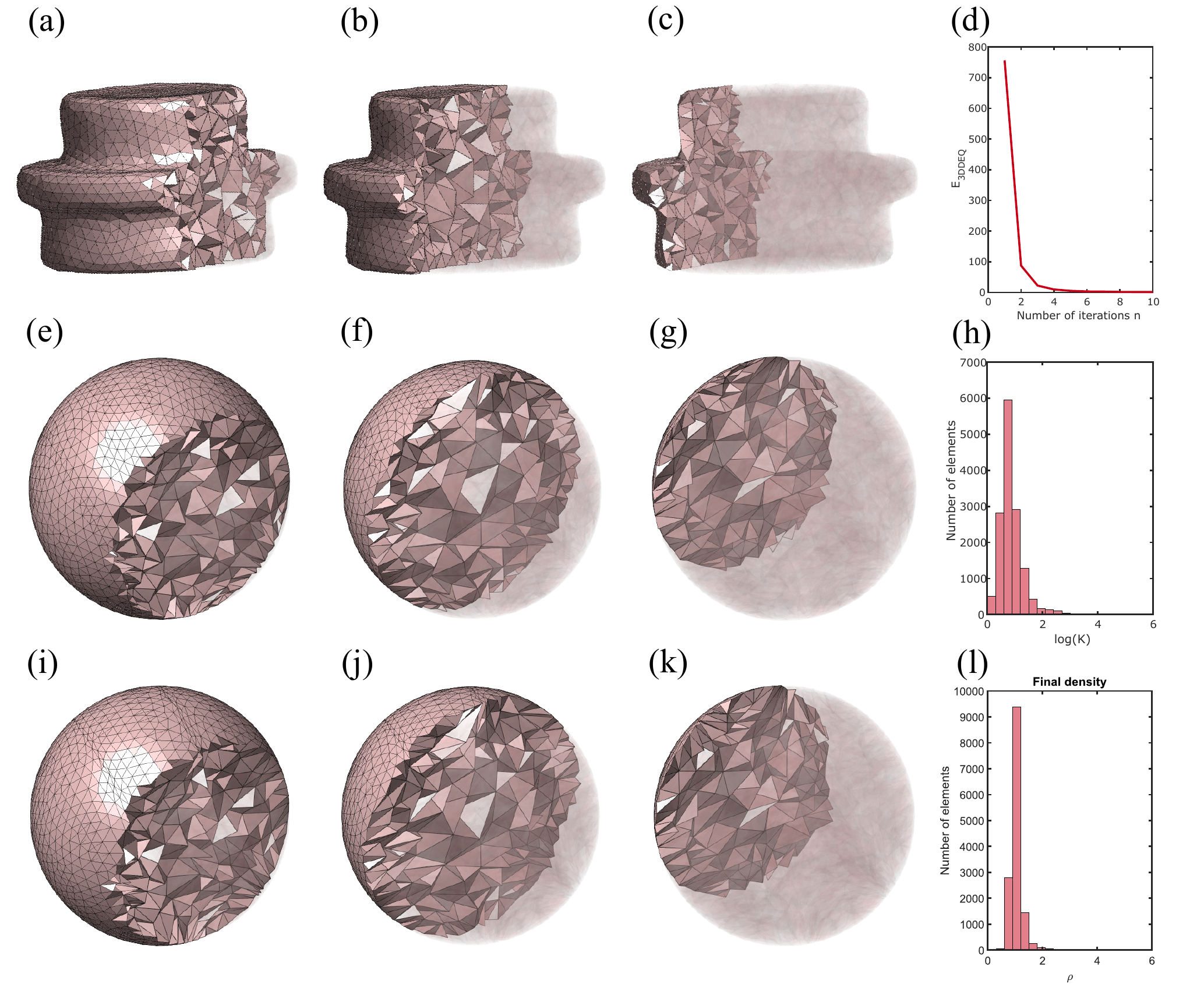}
    \caption{\textbf{3D volume-preserving quasiconformal map of Dual-Axis Guide Sleeve.} Top row (left to right): the original manifold with different slices, and the histogram of the energy $E_{\text{3DDEQ}}$. Middle row (left to right): the initial solid ball with different slices, and the histogram for the geometric coefficient $\overline{K}$. Bottom row (left to right): the result solid ball with different slices, and the histogram for the final density $\rho$.}
    \label{fig:deq_front}
\end{figure}

Fig.~\ref{fig:deq_igea} and~\ref{fig:deq_front} show two examples of mapping general solid 3D manifolds utilizing our proposed 3DDEQ method. In Fig.~\ref{fig:deq_igea}, we map the Igea module to a solid ball while shrinking a certain interior red domain. The population of Igea is the same as Fig.~\ref{fig:deq_igea}. The histograms in Fig.~\ref{fig:deq_igea}(d),(h), and (l) reflect that the result map converges efficiently with lower density and geometric distortions. 

Fig.~\ref{fig:deq_front} presents the bijective volume-preserving quasiconformal result of the Dual-Axis Guide Sleeve. The population is defined as the volume of each tetrahedral element in the original 3D Dual-Axis Guide Sleeve. Initial and final mappings are given in Fig.~\ref{fig:deq_front}(e)-(g) and (i)-(k). The energy histogram in Fig.~\ref{fig:deq_front}(d) indicates that the energy $E_{\text{3DDEQ}}$ converges rapidly throughout the iterations. The final density in Fig.~\ref{fig:deq_front}(l) concentrates at 1, demonstrating that the final result effectively preserves the volume. Additionally, the histogram of the geometric coefficient in Fig.~\ref{fig:deq_front}(h) shows that the geometric distortions are small as well.

\begin{figure}[t]
   \centering
   \includegraphics[width=0.9\textwidth]{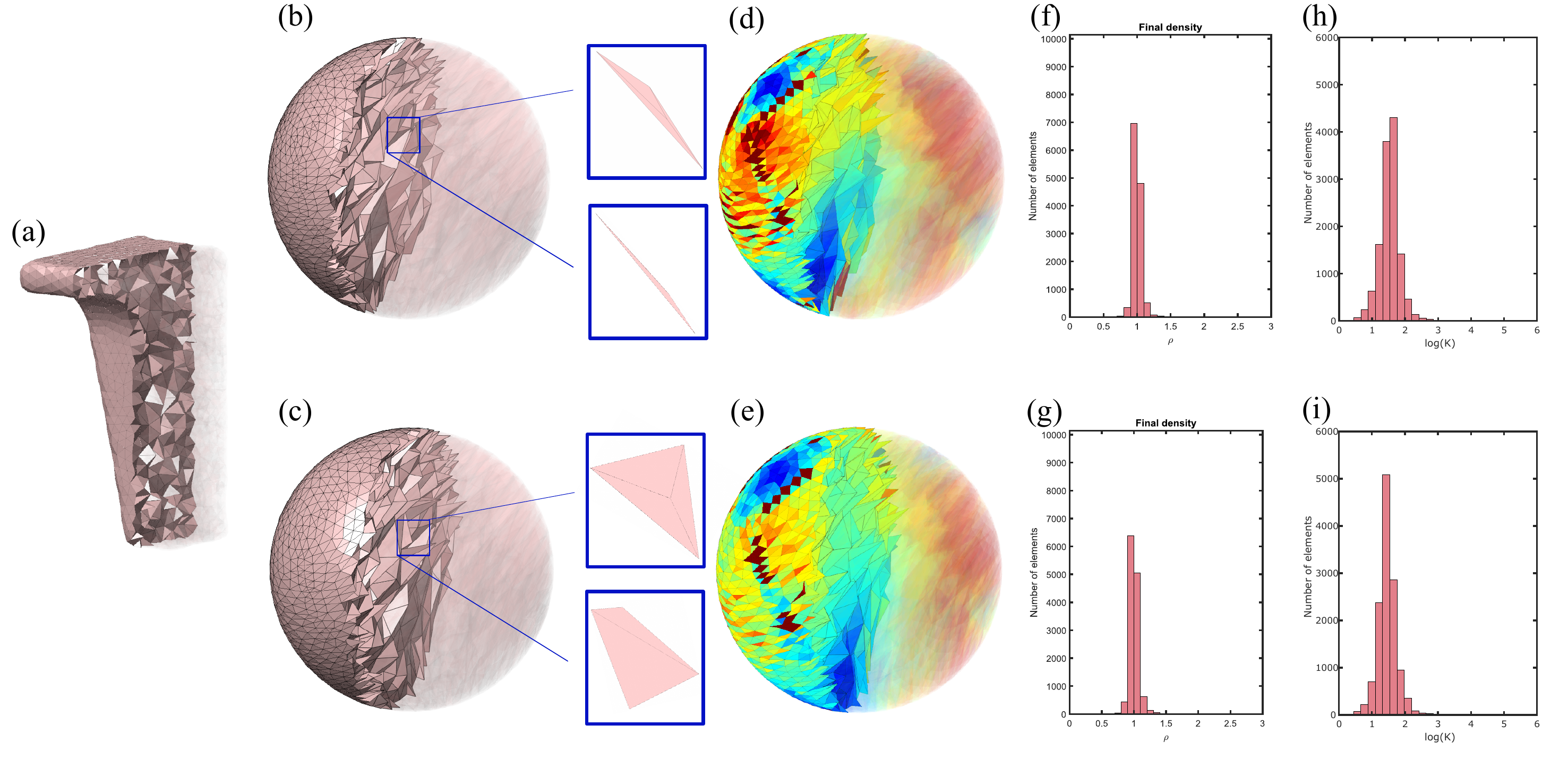}
    \caption{\textbf{Comparison between 3DDEM and 3DDEQ.} (a) The T-Flex Dynamic module. (b) The 3DDEM result with a zoom-in showing the presence of the tetrahedron with large geometric distortions. (c) The 3DDEQ result with a zoom-in of the same tetrahedron. (d) The 3DDEM result colored by the geometric coefficient. (d) The 3DDEQ result colored by the geometric coefficient. (f) The final density histogram of 3DDEM. (g) The final density histogram of 3DDEQ. (h) The histogram of $\overline{K}$ of 3DDEM. (i) The histogram of $\overline{K}$ of 3DDEQ.}
    \label{fig:dem_deq_comparison}
\end{figure}

In Fig.~\ref{fig:dem_deq_comparison}, we compare the proposed 3DDEM and 3DDEQ methods. We consider parameterizing the T-Flex Dynamic module using both methods. The populations are the volume of each tetrahedral element in the given module. As mentioned in the previous section, both methods preserve bijectivity during the iterations. However, the 3DDEQ method considers both density and geometric distortions, while the 3DDEM method focuses solely on density. As shown in the 3DDEM result in Fig.~\ref{fig:dem_deq_comparison}(b), the 3DDEM method may lead to irregular tetrahedrons. By contrast, the proposed 3DDEQ method can make the tetrahedrons more regular as shown in Fig.~\ref{fig:dem_deq_comparison}(c). To show more details, we color the 3DDEM and 3DDEQ results by their corresponding geometric coefficients(Fig.~\ref{fig:dem_deq_comparison}(d)-(e)). It is evident that there are several red domains (indicating high geometric distortions) in Fig.~\ref{fig:dem_deq_comparison}(d). In contrast, these red domains almost disappear in Figure~\ref{fig:dem_deq_comparison}(e), demonstrating that geometric distortions have been significantly reduced. The final densities in Fig.~\ref{fig:dem_deq_comparison}(f)-(g) concentrate at 1, which points out that both results achieve the density-equalizing. The geometric histograms are shown in Fig.~\ref{fig:dem_deq_comparison}(h)-(i), from which we can see that the 3DDEQ method has lower geometric distortions during the iteration process.

Table~\ref{tab:DEQ_DEM_comparison} presents a quantitative comparison between 3DDEM and 3DDEQ methods. Table~\ref{tab:DEQ_DEM_comparison} records the number of tetrahedrons, the variance of the final density $\widetilde{\rho}$, and the mean and standard derivatives of the geometric coefficient $K$. As shown in the table, both 3DDEM and 3DDEQ methods can achieve density-equalizing well. Additionally, compared with the 3DDEM method, the 3DDEQ method reduces the geometric distortions efficiently in all experiments. Moreover, as the bijectivity is guaranteed
by both methods, there is no mesh overlap in the mapping results.

\begin{table}[t]
  \caption{\textbf{Comparison between the proposed 3DDEQ method and the 3DDEM method for 3D manifolds.}}\label{tab:DEQ_DEM_comparison}
\begin{center}
\resizebox{\textwidth}{!}{
\begin{tabular}{|c|c|c|c|c|c|c|c|c|c|}
\hline
\multicolumn{1}{|c|}{ \multirow{2}*{\textbf{Manifold}} }& \multicolumn{1}{c|}{ \multirow{2}*{\textbf{\# tetrahedrons}}} & \multicolumn{4}{c|}{\textbf{3DDEQ}}&\multicolumn{4}{c|}{\textbf{3DDEM}} \\
\cline{3-10}
\multicolumn{1}{|c|}{}&\multicolumn{1}{c|}{} & \bf $\text{Var}(\widetilde{\rho})$  & \bf $\text{mean}(K)$ &\bf $\text{sd}(K)$ &\bf \# Overlaps  & \bf $\text{Var}(\widetilde{\rho})$  & \bf $\text{mean}(K)$ &\bf $\text{sd}(K)$  & \bf \# Overlaps\\
\hline
Cube & 13720 & 0.0015 & 2.1809 & 1.8935 & 0  & 0.0016 & 2.2914 & 1.9886 & 0 \\ \hline
Ellipsoid & 7250 & 0.0012 & 1.9945 & 2.6701 & 0 & 0.0014& 2.1459 & 8.0883 & 0    \\ \hline
Igea & 19539 & 0.0006 & 2.0462 & 1.6577 & 0  & 0.0003 & 2.1151 & 5.5377 & 0  \\\hline
Dual-Axis Guide Sleeve & 14422 & 0.0014 & 2.7179 & 2.9334 & 0 & 0.0016 & 3.0478 & 8.6382 & 0 \\\hline
\end{tabular}
}
\end{center}
\end{table}

\section{Applications}\label{sec:applications}
Remeshing is one of the most important applications of parameterization. It utilizes the inverse mapping to generate high-quality meshes based on the parameterization results. Moreover, different parameterization methods can induce different remeshed results. In the last few decades, numerous remeshing methods have been proposed for the 2D case. For instance, Choi, ~\cite{choi2015flash} proposed a spherical conformal remeshing method. Su et al.~\cite{su2019curvature} computed optimal transport for surface remeshing. Later, some point cloud parameterization methods~\cite{lai2025point,choi2016spherical} have been developed and applied for surface remeshing. However, compared with 2D surfaces, few remeshing methods work for 3D manifolds. 

Motivated by this, we apply our proposed method for 3D manifold remeshing. More specifically, given a 3D solid manifold $\mathcal{M}$, we first compute the special parameterization $f:\mathcal{M} \rightarrow \mathbb B$ utilizing the proposed methods. Then we generate a uniform solid mesh on $\mathbb B$. As aforementioned, we obtain the remeshed manifold by using the inverse mapping $f^{-1}$ to map the uniform solid mesh back onto the given manifold $\mathcal{M}$.

It is noteworthy that the remeshing results are highly connected to the parameterization methods. By computing a quasi-conformal parameterization using 3DQC, we can ensure that the tetrahedral elements will be largely regular in the manifold remeshing result. Additionally, the density-equalizing mapping obtained by 3DDEM can guarantee that the mesh density of the remeshing result is highly uniform. For the 3DDEQ, which balances both density and geometric distortions, the remeshing result can achieve both uniformity and regularity.

\begin{figure}[t]
   \centering
   \includegraphics[width=0.6\textwidth]{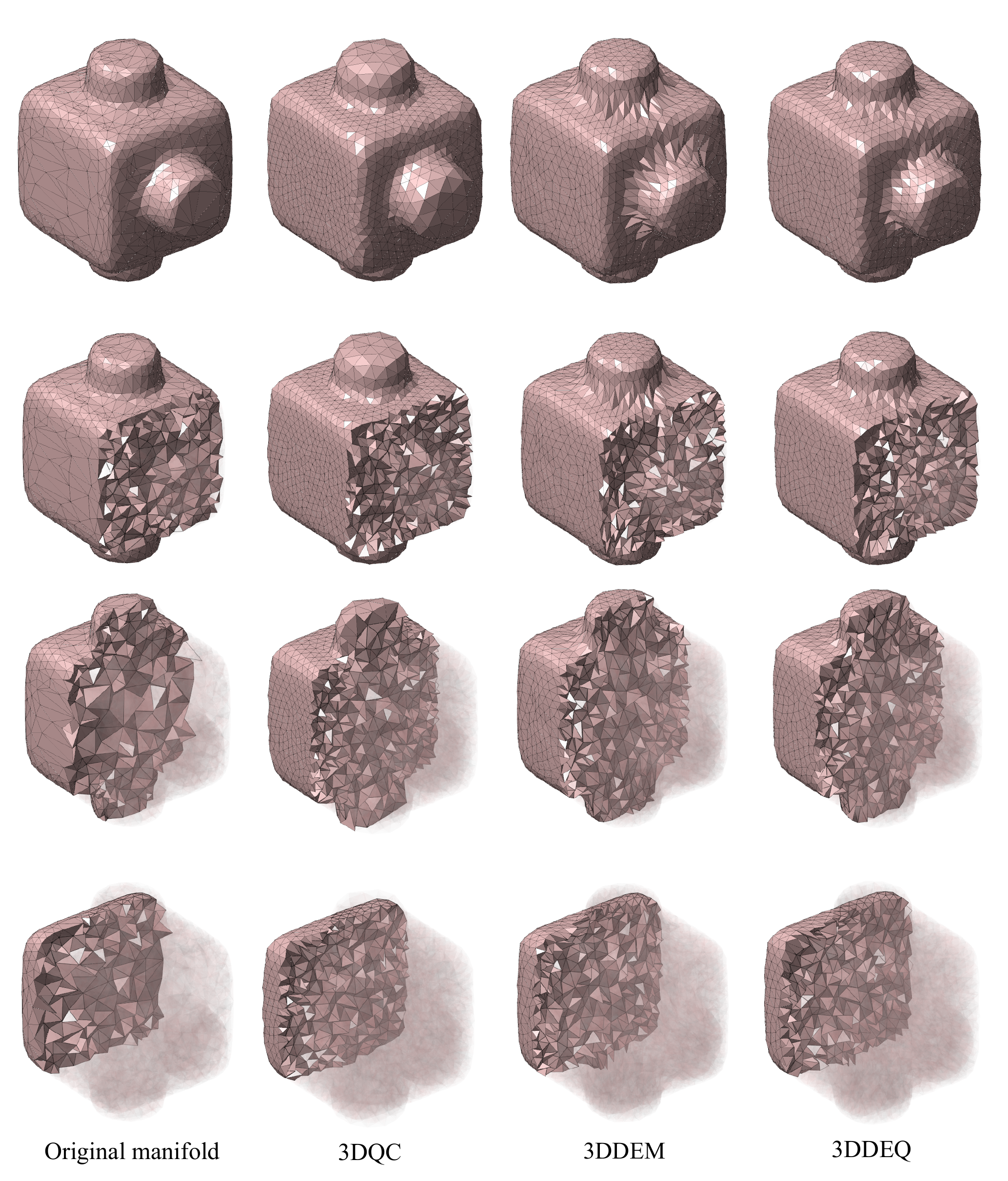}
    \caption{\textbf{Remeshing results for Cross Plug.} The first column shows the original Cross plug module with different slices. The second column shows the remeshing result obtained by 3DQC. The third column shows the remeshing result obtained by 3DDEM. The fourth column shows the remeshing result obtained by 3DDEQ.}
    \label{fig:remesh_result}
\end{figure}

Fig.~\ref{fig:remesh_result} shows the examples of remeshing the Cross Plug module using 3DQC, 3DDEM, and 3DDEQ methods. From the original manifold, we can observe that the initial tetrahedron is highly irregular and non-uniform. The second column of Fig.~\ref{fig:remesh_result} presents the remeshing results of the 3DQC method. Compared with the original manifold, the tetrahedral mesh of this column is much more regular. However, it is still non-uniform, especially on the bulge and corner domains. This can be explained by the fact that the 3DQC method only considers the geometric distortions and does not control the volumetric distortions. By contrast, the 3DDEM method, with the population set proportional to the tetrahedron volume of the original mesh, effectively reduces volumetric distortions. As shown in the third column of Fig.~\ref{fig:remesh_result}, the remeshed output from 3DDEM exhibits a uniform distribution. In the last column of Fig.~\ref{fig:remesh_result}, we present the remeshing result of the 3DDEQ method. It can be observed that we achieve a better remeshing result with a more uniform and regular distribution.

\begin{figure}[t]
   \centering
   \includegraphics[width=0.6\textwidth]{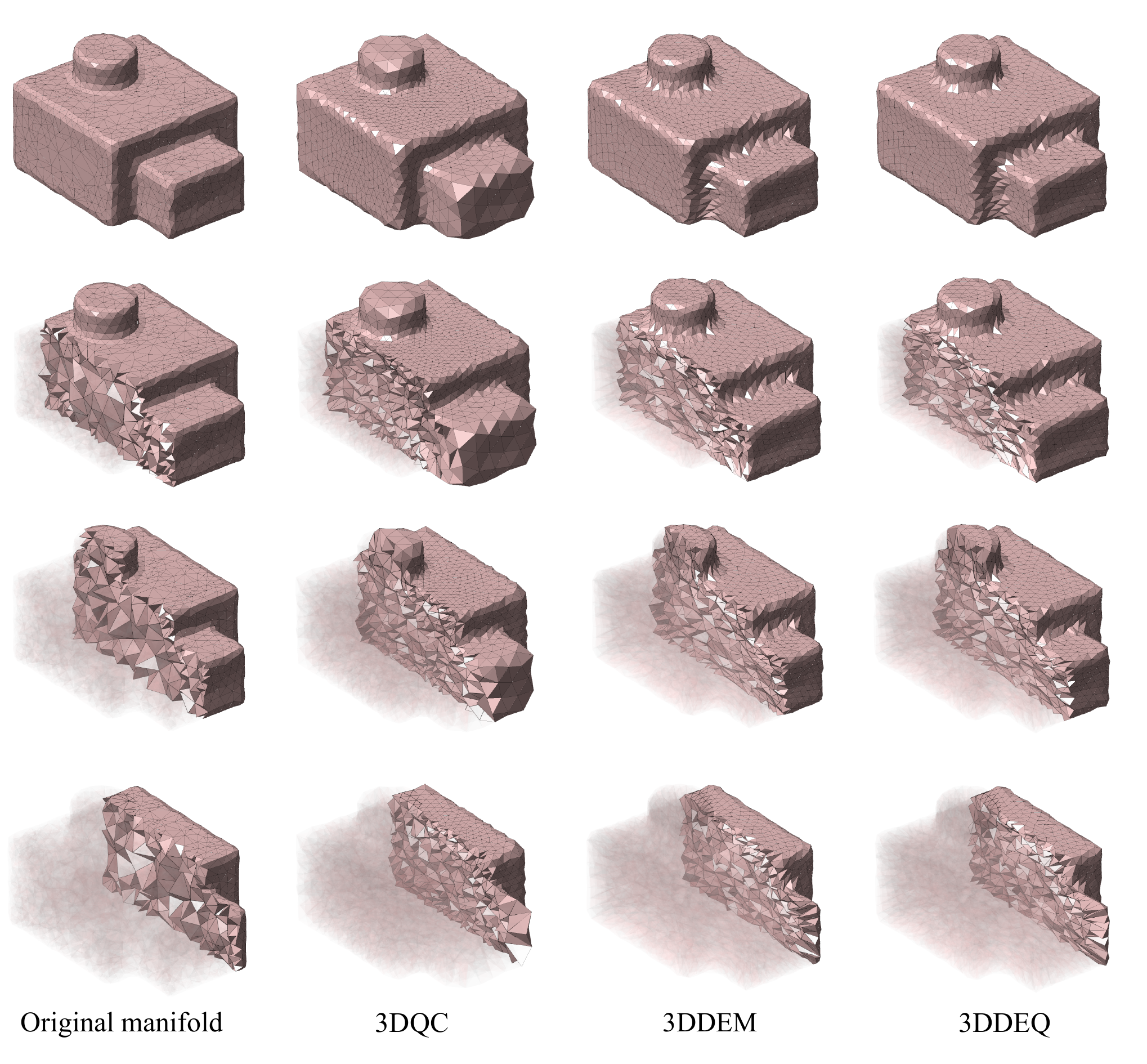}
    \caption{\textbf{Remeshing results for L-shaped connecting element.} The first column shows the original L-shaped connecting element with different slices. The second column shows the remeshing result obtained by 3DQC. The third column shows the remeshing result obtained by 3DDEM. The fourth column shows the remeshing result obtained by 3DDEQ.}
    \label{fig:remesh_result1}
\end{figure}

Fig.~\ref{fig:remesh_result1} shows another example of the L-shaped connecting element. In this case, we can observe that the 3DQC method results in a tetrahedral mesh that is regular in shape but still exhibits non-uniformity, particularly in the areas around the top and front bulge regions. In contrast, the remeshing outcome achieved through the 3DDEM method demonstrates a significantly more uniform distribution of tetrahedral elements. Furthermore, the 3DDEQ method provides an even more optimized result, ensuring that the mesh distribution is both uniform and regular. This combination enhances the overall quality of the remeshed structure, making it suitable for various computational applications.

For a more quantitative analysis, we compute the mean and standard deviations of the 3D quasi-conformal coefficient $K$ for different methods. Moreover, we evaluate the manifold remeshing performances of three approaches by comparing the shape and size variations of the tetrahedral elements in the remeshed meshes. The size variation measure $\delta_{\text{size}}$ is defined as:
\begin{equation}
    \delta_{\text{size}} = \operatorname{std}(T_{\text{Vol}})
\end{equation}
where $T_{\text{Vol}}$ is the volume of each tetrahedron of the remeshed manifold. The parameter $\delta_{\text{size}}$ quantifies the degree of uniformity in the remeshed mesh. Specifically, a smaller value of $\delta_{\text{size}}$ indicates a higher level of uniformity in the mesh structure. The $\delta_{\text{size}}$ equals 0 if and only if all tetrahedral elements are equal in size.

To assess the shape variation, we first define the tetrahedral regularity $R_i$ for each tetrahedron $T_i$ on the remeshed manifold as follows:
\begin{equation}
    R_i = \sum^6_{j = 1}|\frac{e^i_j}{\sum^6_{k = 1}e^i_k} - \frac{1}{6}|,
\end{equation}
where $e^i_k,k = 1,\cdots,6$ represents the length of the six edges of $T_i$. The regularity $R_i = 0$ implies that $T_i$ is an equilateral tetrahedron. Then, the shape variation measure $\delta_{\text{shape}}$ is defined as:
\begin{equation}
    \delta_{\text{shape}} =\underset{i}{ \operatorname{mean}\left(R_i \right)}.
\end{equation}
It is noteworthy that $ \delta_{\text{shape}}$ effectively quantifies the overall regularity of the remeshed manifolds. Specifically, $\delta_{\text{shape}} \geq 0$ for any remeshed mesh, and the equality holds if and only if all tetrahedrons in the remeshed manifold are equilateral.

Table~\ref{tab:remshing_comparison} presents the mean and standard deviation of $K$, as well as the values of $\delta_{\text{shape}}$ and $\delta_{\text{size}}$, for all remeshing results depicted in Figs.~\ref{fig:remesh_result} and \ref{fig:remesh_result1}. Compared to the other two approaches, the 3DQC method exhibits smaller mean and standard deviation values for $K$, indicating that the remeshing results obtained through this method effectively preserve the geometric structure. Additionally, for the 3DQC method, the values of $\delta_{\text{shape}}$ are relatively low, while the values of $\delta_{\text{size}}$ are comparatively high, suggesting that the resulting meshes are highly regular in shape despite variations in size.
In contrast, the values of $\delta_{\text{size}}$ are smaller in density-equalizing approaches, indicating a high degree of uniformity in the remeshing meshes. More specifically, when evaluating both $\delta_{\text{shape}}$ and $\delta_{\text{size}}$ for the two density-equalizing methods, it becomes clear that the remeshing outputs of the 3DDEQ method demonstrate the best overall remeshing performance.

\begin{table}[t]
\small
    \caption{\textbf{Comparison between the surface remeshing results obtained by 3DQC, 3DDEM, and 3DDEQ.} For each method, we record the mean $\text{mean}(K)$ and standard deviation $\text{sd}(K)$ of the 3D quasi-conformal coefficient $K$ of the remeshed manifolds.}\label{tab:remshing_comparison}
    
  \begin{center}
  \begin{tabular}{|c|c|c|c|c|c|} \hline
    \bf Surface & \bf Method & \bf $\text{mean}(K)$  & \bf $\text{sd}(K)$ & \bf $\delta_\text{shape}$ & \bf $\delta_\text{size}$  \\ \hline

    \multirow{3}{*}{Cross Plug} & 3DQC & 1.4960 & 0.6507 & 0.1519 & 0.3590  \\ \cline{2-6}
     & 3DDEM  & 2.7750  & 16.5683 & 0.1909 & 0.2450 \\ \cline{2-6}
    & 3DDEQ  & 1.8802 &  2.1975 & 0.1812 & 0.1981   \\ \hline
    
    \multirow{3}{*}{L-shaped connecting element} & 3DQC  & 1.8399  &  1.3827 & 0.1684 & 0.3615 \\ \cline{2-6}
     & 3DDEM  & 2.8907  & 10.2453 & 0.2220 & 0.2528 \\ \cline{2-6}
    & 3DDEQ  & 2.6626  &  4.2031 & 0.2129 & 0.2250  \\ \hline
  \end{tabular}
  
\end{center}
\end{table}
\section{Discussion}\label{sec:discussion}
In this work, we proposed a novel framework for parameterizing 3D solid manifolds. This framework enables the generation of parameterizations with specific, desirable distortion properties. By minimizing the 3D quasi-conformal coefficient K, we achieve 3D quasi-conformal mappings (3DQC). Minimizing a diffusion term yields bijective 3D density-equalizing mappings (3DDEM). Combining these formulations, we introduce a 3D Density-Equalized Quasi-conformal map (3DDEQ), which optimally balances volumetric density distortion with angular distortion. The experimental results and applications have demonstrated the effectiveness of our proposed methods.

Note that our proposed framework is limited to manifolds with solid ball topology. In the future, we plan to extend the methods for manifolds with our topologies. Moreover, landmark-matching and hybrid registration constraints have not been incorporated into our
framework. Another possible future direction is to extend our methods to facilitate landmark-matching manifold parameterization and manifold registration.

\bibliographystyle{ieeetr}
\bibliography{3DDEQbib.bib}
\end{document}